\documentclass[11pt]{article}

\usepackage{graphicx}
\usepackage{amsmath}
\usepackage{amsfonts}
\usepackage{amssymb}
\usepackage{enumerate}
\usepackage{lscape}
\usepackage{longtable}
\usepackage{color}
\usepackage{url}

\newtheorem{theorem}{Theorem}

\newtheorem{algorithm}[theorem]{Algorithm}

\newtheorem{claim}{Claim}[section]

\newtheorem{conjecture}[theorem]{Conjecture}

\newtheorem{definition}[theorem]{Definition}
\newtheorem{lemma}[theorem]{Lemma}

\newtheorem{proposition}[theorem]{Proposition}

\newenvironment{proof}[1][Proof]{\textbf{#1.} }{\ \rule{0.5em}{0.5em} \vspace{1ex}}

\def\EE{\mathbb{E}}
\newcommand{\e}{\mathbb{E}}

\newcommand{\qed}{\hfill \square}

\newcommand{\OOO}{\mathcal{O}}
\newcommand{\UUU}{\mathcal{U}}

\newcommand{\PPP}{\mathcal{P}}
\newcommand{\NNN}{\mathcal{N}}
\newcommand{\DDD}{\mathcal{D}}

\newcommand{\tr}{\operatorname{tr}}

\renewcommand{\iota}{{l}}
\newcommand{\card}{{\rm card}}
\newcommand{\argmin}{\operatorname{argmin}}

\newcommand{\RR}{\mathbb{R}}
\newcommand{\CC}{\mathbb{C}}
\newcommand{\KK}{\mathbb{K}}
\newcommand{\NN}{\mathbb{N}}

\newcommand{\Prob}{\mathrm{Prob}}
\newcommand{\sign}{\mathrm{sign}}

\begin{document}

\title{Approximating the Little Grothendieck Problem over the Orthogonal and Unitary Groups}
\author{
Afonso S. Bandeira\thanks{Department of Mathematics, Massachusetts Institute of Technology, Cambridge, Masachusetts 02142, USA ({\tt bandeira@mit.edu}). Most of this work was done while ASB was at Princeton University. 
}
\and
Christopher Kennedy\thanks{Department of Mathematics, The University of Texas at Austin, Austin, Texas 78712, USA ({\tt ckennedy@math.utexas.edu}).
}
\and
Amit Singer%
\thanks{Department of Mathematics and PACM, Princeton University, Princeton, New Jersey 08544, USA ({\tt amits@math.princeton.edu}).
}
}

\maketitle
{\small


\footnotesep=0.4cm

\begin{abstract}
The little Grothendieck problem consists of maximizing $\sum_{ij}C_{ij}x_ix_j$ for a positive semidefinite matrix $C$, over binary variables $x_i\in\{\pm1\}$.
In this paper we focus on a natural generalization of this problem, the little Grothendieck problem over the orthogonal group. Given $C\in\RR^{dn\times dn}$ a positive semidefinite matrix, the objective is to maximize $\sum_{ij}\tr\left(C^T_{ij}O_iO_j^T\right)$ restricting $O_i$ to take values in the group of orthogonal matrices $\OOO_d$, where $C_{ij}$ denotes the $(ij)$-th $d\times d$ block of $C$.

We propose an approximation algorithm, which we refer to as \emph{Orthogonal-Cut}, to solve the little Grothendieck problem over the group of orthogonal matrices $\OOO_d$ and show a constant approximation ratio. Our method is based on semidefinite programming. 
For a given $d\geq 1$, 
we show a constant approximation ratio of $\alpha_{\RR}(d)^2$, where $\alpha_{\RR}(d)$ is the expected average singular value of a $d\times d$ matrix with random Gaussian $\NNN\left(0,\frac1d\right)$ i.i.d. entries. For $d=1$ we recover the known $\alpha_{\RR}(1)^2=2/\pi$ approximation guarantee for the classical little Grothendieck problem. Our algorithm and analysis naturally extends to the complex valued case also providing a constant approximation ratio for the analogous little Grothendieck problem over the Unitary Group $\UUU_d$.

Orthogonal-Cut also serves as an approximation algorithm for several applications, including the Procrustes problem where it improves over the best previously known approximation ratio of~$\frac1{2\sqrt{2}}$.
The little Grothendieck problem falls under the larger class of problems approximated by a recent algorithm proposed in the context of the non-commutative Grothendieck inequality. Nonetheless, our approach is simpler and provides better approximation with matching integrality gaps. 

Finally, we also provide an improved approximation algorithm for the more general little Grothendieck problem over the orthogonal (or unitary) group with rank constraints, recovering, when $d=1$, the sharp, known ratios.

\end{abstract}


\begin{center}
\end{center}

}

\pagebreak


\section{Introduction}


The little Grothendieck problem~\cite{NAlon_ANaor_2006} in combinatorial optimization is written as
\begin{equation}\label{littleGrothendieck}
 \max_{x_i\in \{\pm1\}}\sum_{i=1}^n\sum_{j=1}^nC_{ij}x_ix_j,
\end{equation}
where $C$ is a $n\times n$ positive semidefinite matrix real matrix.

Problem (\ref{littleGrothendieck}) is known to be NP-hard. In fact, if $C$ is a Laplacian matrix of a graph then (\ref{littleGrothendieck}) is equivalent to the Max-Cut problem.
In a seminal paper in the context of the Max-Cut problem, Goemans and Williamson~\cite{MXGoemans_DPWilliamson_1995} provide a semidefinite relaxation for (\ref{littleGrothendieck}):
\begin{equation}
\sup_{m\in \NN}\max_{\substack{ X_i\in\RR^{m}  \\ \|X_i\|^2=1}} \sum_{i=1}^n\sum_{j=1}^n C_{ij}X_i^TX_j.
\label{eq_OCSDPv1}
\end{equation}
It is clear that in (\ref{eq_OCSDPv1}), one can take $m=n$. Furthermore, (\ref{eq_OCSDPv1}) is equivalent to a semidefinite program and can be solved, to arbitrary precision, in polynomial time~\cite{LVanderberghe_SBoyd_1996}. In the same paper~\cite{MXGoemans_DPWilliamson_1995} it is shown that a simple rounding technique is guaranteed to produce a solution whose objective value is, in expectation, at least a multiplicative factor $\frac2\pi\min_{0\leq \theta \leq \pi} \frac{\theta}{1-\cos\theta} \approx 0.878$ of the optimum.

A few years later, Nesterov~\cite{Nesterov_quadprogram1} showed an approximation ratio of $\frac2\pi$ for the general case of an arbitrary positive semidefinite $C\succeq 0$ using the same relaxation as~\cite{MXGoemans_DPWilliamson_1995}. This implies, in particular, that the value of (\ref{littleGrothendieck}) can never be smaller than $\frac2{\pi}$ times the value of (\ref{eq_OCSDPv1}). Interestingly, such an inequality was already known from the influential work of Grothendieck on norms of tensor products of Banach spaces~\cite{Grothendieck_GT} (see~\cite{Pisier_GT} for a survey on this).

Several more applications have since been found for the Grothendieck problem (and variants), and its semidefinite relaxation. Alon and Naor~\cite{NAlon_ANaor_2006} showed applications to estimating the cut-norm of a matrix; Ben-Tal and Nemirovski~\cite{Ben-tal_Nemirovski_02} showed applications to control theory; Briet, Buhrman, and Toner~\cite{Briet09generalizedGrothendieckXOR} explored connections with quantum non-locality; and many more (see~\cite{Alon_etal_generalGrothendieck}).

In this paper, we focus on a natural generalization of problem (\ref{littleGrothendieck}), the little Grothendieck problem over the orthogonal group, where the variables are now elements of the orthogonal group $\OOO_d$, instead of $\{\pm1\}$. More precisely, given $C\in\RR^{dn\times dn}$ a positive semidefinite matrix, we consider the problem
\begin{equation}
\max_{O_1,.. .,O_n \in \OOO_d} \sum_{i=1}^n\sum_{j=1}^n\tr\left(C_{ij}^TO_iO_j^T\right),
\label{eq_QFOG}
\end{equation}
where $C_{ij}$ denotes the $(i,j)$-th $d\times d$ block of $C$, and $\OOO_d$ is the group of $d\times d$ orthogonal matrices (i.e., $O\in  \OOO_d$ if and only if $OO^T=O^TO=I_{d\times d}$).

We will also consider the unitary group variant, where the variables are now elements of the unitary group $\UUU_d$ (i.e., $U\in \UUU_d$ if and only if $UU^H=U^HU=I_{d\times d}$). More precisely, given $C\in\CC^{dn\times dn}$ a complex valued positive semidefinite matrix, we consider the problem
\begin{equation}
\max_{U_1,.. .,U_n \in \UUU_d} \sum_{i=1}^n\sum_{j=1}^n\tr\left(C_{ij}^HU_iU_j^H\right).
\label{eq_QFUG}
\end{equation}
Since $C$ is Hermitian positive semidefinite, the value of the objective function in (\ref{eq_QFUG}) is always real. Note also that when $d=1$, (\ref{eq_QFOG}) reduces to (\ref{littleGrothendieck}). Also, since $\UUU_1$ is the multiplicative group of the complex numbers with unit norm, (\ref{eq_QFUG}) recovers the classical complex case of the little Grothendieck problem. In fact, the work of Nesterov was extended~\cite{So_Zhang_Ye_QuadSDPrelax} to the complex plane (corresponding to $\UUU_1$, or equivalently, the special orthogonal group $S\OOO_2$) with an approximation ratio of $\frac{\pi}4$ for $C\succeq 0$. As we will see later, the analysis of our algorithm shares many ideas with the proofs of both~\cite{Nesterov_quadprogram1} and~\cite{So_Zhang_Ye_QuadSDPrelax} and recovers both results.

As we will see in Section~\ref{section:applications}, several problems can be written in the forms~(\ref{eq_QFOG}) and~(\ref{eq_QFUG}), such as the Procrustes problem~\cite{Schonenmann66_Procrustes,Nemirovski_sumsProcrustes,So_sumsProcrustes} 
and Global Registration~\cite{Chaudhury_etal_GlobalRegistration}. Moreover, the approximation ratio we obtain for (\ref{eq_QFOG}) and (\ref{eq_QFUG}) translates into the same approximation ratio for these applications, improving over the best previously known approximation ratio of $\frac1{2\sqrt{2}}$ in the real case and $\frac12$ in the complex case, given by~\cite{Naor_etal_NCGI} for these problems.

Problem (\ref{eq_QFOG}) belongs to a wider class of problems considered by Nemirovski~\cite{Nemirovski_sumsProcrustes} called {QO-OC} (Quadratic Optimization under Orthogonality Constraints), which itself is a subclass of QC-QP (Quadratically Constrainted Quadratic Programs). Please refer to Section~\ref{section:applications} for a more detailed comparison with the results of Nemirovski~\cite{Nemirovski_sumsProcrustes}. 
More recently, Naor et al.~\cite{Naor_etal_NCGI} propose an efficient rounding for the non commutative Grothendieck inequality that provides an approximation algorithm for a vast set of problems involving orthogonality constraints, including problems of the form of (\ref{eq_QFOG}) and (\ref{eq_QFUG}).  We refer to Section~\ref{sec:CompareNRV} for a comparison between this approach and ours.


Similarly to (\ref{eq_OCSDPv1}) we formulate a semidefinite relaxation we name the \emph{Orthogonal-Cut} SDP:
\begin{equation}
\sup_{m\in \NN}\max_{\substack{ X_iX_i^T=I_{d\times d} \\ X_i\in\RR^{d\times m} }} \sum_{i=1}^n\sum_{j=1}^n\tr\left(C_{ij}^TX_iX_j^T\right).
\label{eq_OCSDPv}
\end{equation}
Analogously, in the unitary case, we consider the relaxation
\begin{equation}
\sup_{m\in \NN}\max_{\substack{ Y_iY_i^H=I_{d\times d} \\ Y_i\in\CC^{d\times m} }} \sum_{i=1}^n\sum_{j=1}^n\tr\left(C_{ij}^HY_iY_j^H\right).
\label{eq_UCSDPv}
\end{equation}
Since $C$ is Hermitian positive semidefinite, the value of the objective function in (\ref{eq_UCSDPv}) is guaranteed to be real. Note also that we can take $m=dn$ as the Gram matrix $[X_iX_j^T]_{i,j}$ does not have a rank constraint for this value of $m$. In fact, both problems (\ref{eq_OCSDPv}) and (\ref{eq_UCSDPv}) are equivalent to the semidefinite program
\begin{equation}
\max_{\substack{ G\in\KK^{dn\times dn} \\ G_{ii}=I_{d\times d} ,\ G\succeq 0  }} \tr(CG),
\label{eq_OCSDPm}
\end{equation}
for $\KK$ respectively $\RR$ and $\CC$, and so can be solved, up to arbitrary precision, in polynomial time\footnote{We also note that these semidefinite programs satisfy Slater's condition as the identity matrix is a feasible point. This ensures strong duality, which can be exploited by many semidefinite programming solvers.}~\cite{LVanderberghe_SBoyd_1996}. At first glance, one could think of problem (\ref{eq_OCSDPv}) as having $d^2n$ variables and that we would have to take $m=d^2n$ for (\ref{eq_OCSDPv}) to be tractable (in fact, this is the size of the SDP considered by Nemirovski~\cite{Nemirovski_sumsProcrustes}). The savings in size (corresponding to number of variables) of our proposed SDP relaxation come from the group structure of $\OOO_d$ (or $\UUU_d$).

One of the main contributions of this paper is showing that Algorithm~\ref{algorithm:main} (Section~\ref{section:algorithm}) gives a constant factor approximation to (\ref{eq_QFOG}), and its unitary analog (\ref{eq_QFUG}), with an optimal approximation ratio for our relaxation (Section~\ref{section:integralitygap}). It consists of a simple generalization of the rounding in~\cite{MXGoemans_DPWilliamson_1995} applied to (\ref{eq_OCSDPv}), or (\ref{eq_QFUG}).
\begin{theorem}\label{lemma_maintheorem}
Let $C\succeq 0$ and real. Let $V_1,\dots,V_n\in \OOO_d$ be the (random) output of the orthogonal version of Algorithm~\ref{algorithm:main}. Then
\[
\EE \left[ \sum_{i=1}^n\sum_{j=1}^n\tr\left(C_{ij}^TV_iV_j^T\right) \right] \geq \alpha_{\RR}(d)^2 \max_{O_1,.. .,O_n \in \OOO_d} \sum_{i=1}^n\sum_{j=1}^n\tr\left(C_{ij}^TO_iO_j^T\right),
\]
where $\alpha_{\RR}(d)$ is the constant defined below.

Analogously, in the unitary case, if $W_1,\dots,W_n\in \UUU_d$ are the (random) output of the unitary version of Algorithm~\ref{algorithm:main}, then for $C\succeq 0$ and complex,
\[
\EE \left[ \sum_{i=1}^n\sum_{j=1}^n\tr\left(C_{ij}^HW_iW_j^H\right) \right] \geq \alpha_{\CC}(d)^2 \max_{U_1,.. .,U_n \in \UUU_d} \sum_{i=1}^n\sum_{j=1}^n\tr\left(C_{ij}^HU_iU_j^H\right),
\]
where $\alpha_{\CC}(d)$ is defined below.
\end{theorem}

\begin{definition}\label{def:alphad}
 Let $G_{\RR} \in \RR^{d\times d}$ and $G_{\CC} \in \CC^{d\times d}$ be, respectively, a Gaussian random matrix with i.i.d real valued entries $\NNN\left(0,d^{-1}\right)$ and a Gaussian random matrix with i.i.d complex valued entries $\NNN\left(0,d^{-1}\right)$.  We define
\[
\alpha_{\RR}(d) := \e\left[ \frac1d \sum_{j=1}^d \sigma_j(G_{\RR})\right] \text{ and } \alpha_{\CC}(d) := \e\left[ \frac1d \sum_{j=1}^d \sigma_j(G_{\CC})\right],
\]
where $\sigma_j(G)$ is the $j$th singular value of $G$.
\end{definition}

Although we do not have a complete understanding of the behavior of $\alpha_{\RR}(d)$ and $\alpha_{\CC}(d)$ as functions of $d$, we can, for each $d$ separately, compute a closed form expression (see Section~\ref{section:onalphad}). For $d=1$ we recover the sharp $\alpha_{\RR}(1)^2=\frac2\pi$ and $\alpha_{\CC}(1)^2=\frac{\pi}4$ results of, respectively, Nesterov~\cite{Nesterov_quadprogram1} and So et al.~\cite{So_Zhang_Ye_QuadSDPrelax}. One can also show that $\lim_{d\to\infty}\alpha_{\KK}(d)^2 = \left(\frac{8}{3\pi}\right)^2$, for both $\KK=\RR$ and $\KK=\CC$. Curiously,
$$\alpha_{\RR}(1)^2 =  \frac2{\pi} < \left(\frac{8}{3\pi}\right)^2 <  \frac{\pi}4 = \alpha_{\CC}(1)^2.$$ Our computations strongly suggest that $\alpha_{\RR}(d)$ is monotonically increasing while its complex analog $\alpha_{\CC}(d)$ is monotonically decreasing. We find the fact that the approximation ratio seems to get, as the dimension increases, better in the real case and worse in the complex case quite intriguing. One might naively think that the problem for a specific $d$ can be formulated as a degenerate problem for a larger $d$, however this does not seem to be true, as evidenced by the fact that $\alpha_{\RR}^2(d)$ is increasing. Another interesting point is that $\alpha_{\RR}(2) \neq \alpha_{\CC}(1)$ which suggests that the little Grothendieck problem over $\OOO_2$ is quite different from the analog in $\UUU_1$ (which is isomorphic to $S\OOO_2$).
Unfortunately, we were unable to provide a proof for the monotonicity of $\alpha_{\KK}(d)$ (Conjecture~\ref{conjecture:alphar}).
Nevertheless, we can show lower bounds for both $\alpha_{\RR}^2(d)$ and $\alpha_{\CC}^2(d)$ that have the right asymptotics (see Section~\ref{section:onalphad}). In particular, we can show that our approximation ratios are uniformly bounded below by the approximation ratio given in~\cite{Naor_etal_NCGI}.

In some applications, such as the Common Lines problem~\cite{ASinger_YShkolnisky_commonlines} (see Section~\ref{Section:HigherRank}), one is interested in a more general version of (\ref{eq_QFOG}) where the variables take values in the Stiefel manifold $\OOO_{(d,r)}$, the set of matrices $O\in\RR^{d\times r}$ such that $OO^T = I_{d\times d}$. This motivates considering a generalized version of (\ref{eq_QFOG}) formulated as, for $r \geq d$,
\begin{equation}
\max_{O_1,.. .,O_n \in \OOO_{(d,r)}} \sum_{i=1}^n\sum_{j=1}^n\tr\left(C_{ij}^TO_iO_j^T\right),
\label{eq_QFOGrINTRO}
\end{equation}
for $C\succeq0$. The special case $d=1$ was formulated and studied in~\cite{Briet09generalizedGrothendieckXOR} and~\cite{Jop_littleGTrank} in the context of quantum non-locality and quantum XOR games. Note that in the special case $r=nd$, (\ref{eq_QFOGrINTRO}) reduces to (\ref{eq_OCSDPv}) and is equivalent to a semidefinite program.

We propose an adaption of Algorithm~\ref{algorithm:main}, Algorithm~\ref{algorithm:rankr}, and show an approximation ratio of $\alpha_{\RR}(d,r)^2$, where $\alpha_{\RR}(d,r)$ is also defined as the average singular value of a Gaussian matrix (see Section~\ref{Section:HigherRank}). For $d=1$ we recover the sharp results of Briet el al.~\cite{Jop_littleGTrank} giving a simple interpretation for the approximation ratios, as $\alpha(1,r)$ is simply the mean of a normalized chi-distribution with $r$ degrees of freedom. As before, the techniques are easily extended to the complex valued case.

In order to understand the optimality of the approximation ratios $\alpha_{\RR}(d)^2$ and $\alpha_{\CC}(d)^2$ we provide an integrality gap for the relaxations (\ref{eq_OCSDPv}) and (\ref{eq_UCSDPv}) that matches these ratios, showing that they are tight. Our construction of an instance having this gap is an adaption of the classical construction for the $d=1$ case (see, e.g.,~\cite{NAlon_ANaor_2006}). As it will become clear later (see Section~\ref{section:integralitygap}), there is an extra difficulty in the $d>1$ orthogonal case which can be dealt with using the Lowner-Heinz Theorem on operator convexity (see Theorem~\ref{alphastaralpha} and the notes~\cite{Carlen_QuantumEntropy}).

Besides the monotonicity of $\alpha_{\KK}^2(d)$ (Conjecture \ref{conjecture:alphar}), there are several interesting questions raised from this work, including the hardness of approximation of the problems considered in this paper (see Section~\ref{section:futurework} for a discussion on these and other directions for future work).

%
%
%
%

\textbf{Organization of the paper:} The paper is organized as follows. In Section~\ref{section:algorithm} below we present the approximation algorithm for (\ref{eq_QFOG}) and (\ref{eq_QFUG}). In Section~\ref{sec:CompareNRV}, we compare our results with the ones in~\cite{Naor_etal_NCGI}. We then describe a few applications in Section~\ref{section:applications} and show the analysis for the approximation ratio guarantee in Section~\ref{Section:analysisalgorithm}. In Section~\ref{section:onalphad} we analyze the value of the approximation ratio constants. Section~\ref{Section:HigherRank} is devoted to a more general, rank constrained, version of (\ref{eq_QFUG}). We give an integrality gap for our relaxation in Section~\ref{section:integralitygap} and discuss open problems and future work in Section~\ref{section:futurework}. Finally, we present supporting technical results in the Appendix.

\subsection{Algorithm}\label{section:algorithm}

We now present the (randomized) approximation algorithm we propose to solve (\ref{eq_QFOG}) and (\ref{eq_QFUG}).

\begin{algorithm}\label{algorithm:main}
 Compute $X_1,\dots,X_n\in \RR^{d\times nd}$ (or $Y_1,\dots,Y_n\in \CC^{d\times nd}$) a solution to (\ref{eq_OCSDPv}) (or (\ref{eq_UCSDPv})).
Let $R$ be a $nd\times d$ Gaussian random matrix whose entries are real (or complex) i.i.d. $\NNN(0,\frac1d)$.
The approximate solution for (\ref{eq_QFOG}) (or (\ref{eq_QFUG})) is now computed as
\[
V_i = \PPP(X_i R),
\]
where $\PPP(X)=\argmin_{Z\in \OOO_d}\|Z-X\|_F$ (or $\PPP(Y)=\argmin_{Z\in \UUU_d}\|Z-Y\|_F$), for any $X\in\RR^{d\times d}$ (or $Y\in\CC^{d\times d}$) and $||X||_F = \sqrt{\tr\left(X X^T\right)}$($||Y||_F = \sqrt{\tr\left(Y Y^H\right)}$) is the Frobenius norm.
\end{algorithm}

Note that (\ref{eq_OCSDPv}) and (\ref{eq_UCSDPv}) can be solved with arbitrary precision in polynomial time~\cite{LVanderberghe_SBoyd_1996} as they are equivalent to a semidefinite program (followed by a Cholesky decomposition) with a, respectively real and complex valued, matrix variable of size $dn\times dn$, and $d^2n$ linear constraints. In fact, this semidefinite program has a very similar structure to the classical Max-Cut SDP. This may allow one to adapt specific methods designed to solve the Max-Cut SDP such as, for example, the row-by-row method~\cite{RBR-WenGoldfarbMaScheinberg-2009} (see Section 2.4 of~\cite{Bandeira_PhdThesis}).

Moreover, given $X$ a $d\times d$ matrix (real or complex), the polar component $\PPP(X)$  is the orthogonal (or unitary) matrix part of the polar decomposition, that can be easily computed via the singular value decomposition of $X=U\Sigma V^H$ as $\PPP(X) = UV^H$ (see \cite{Fan_polardecomposition,Keller_polardecomposition,Higham_PolarDecomposition}), rendering Algorithm~\ref{algorithm:main} efficient. The polar component $\PPP(X) = UV^H$ is the analog in high dimensions of the sign in $\OOO_1$ and the angle in $\UUU_1$ and can also be written as $\PPP(X) = X\left( X^HX \right)^{-\frac12}$.  


\subsection{Relation to non-commutative Grothendieck inequality}\label{sec:CompareNRV}

The approximation algorithm proposed in~\cite{Naor_etal_NCGI} can also be used to approximate problems (\ref{eq_QFOG}) and (\ref{eq_QFUG}). In fact, the method in~\cite{Naor_etal_NCGI} deals with problems of the form
\begin{equation}\label{NCGI_comb}
\sup_{X,Y\in \OOO_N} \sum_{pqkl}M_{pqkl}X_{pq}Y_{kl},
\end{equation}
where $M$ is a $N\times N\times N\times N$ real valued 4-tensor.

Problem (\ref{eq_QFOG}) can be encoded in the form of (\ref{NCGI_comb}) by taking $N=dn$ and having the $d\times d$ block of $M$, obtained by having the first two indices range from $(i-1)d+1$ to $id$ and the last two from $(j-1)d+1$ to $jd$, equal to $C_{ij}$, and the rest of the tensor equal to zero~\cite{Naor_etal_NCGI}. More explicitly, the nonzero entries of $M$ are given by $M_{(i-1)d+r,(i-1)d+r,(j-1)d+s,(j-1)d+s} = \left[C_{ij}\right]_{rs}$, for each $i,j$ and $r,s=1,\dots,d$. Since $C$ is positive semidefinite, the supremum in (\ref{NCGI_comb}) is attained at a pair $(X,Y)$ such that $X=Y$.

In order to describe the relaxation one needs to first define the space of vector-valued orthogonal matrices $\OOO_N(m) = \{ X\in\RR^{N\times N\times m}: XX^T = X^TX = I_{N\times N}  \}$ where $XX^T$ and $X^TX$ are $N\times N$ matrices defined as $\left(XX^T\right)_{pq} = \sum_{k=1}^N\sum_{r=1}^m X_{pkr}X_{qkr} \text{ and }  \left(X^TX\right)_{pq} = \sum_{k=1}^N\sum_{r=1}^m X_{kpr}X_{kqr}$.

The relaxation proposed in~\cite{Naor_etal_NCGI} (which is equivalent to our relaxation when $M$ is specified as above) is given by
\begin{equation}\label{NCGI_sdp}
\sup_{m\in\mathbb{N}}\ \sup_{U,V\in \OOO_N(m)}\ \sum_{pqkl}M_{pqkl}U_{pq}V_{kl},
\end{equation}
and there exists a rounding procedure~\cite{Naor_etal_NCGI} that achieves an approximation ratio of $\frac1{2\sqrt{2}}$. Analogously, in the unitary case, the relaxation is essentially the same and the approximation ratio is $\frac12$. We can show (see Section \ref{section:onalphad}) that the approximation ratios we obtain are larger than these for all $d\geq 1$. Interestingly, the approximation ratio of $\frac12$, for the complex case in~\cite{Naor_etal_NCGI}, is tight in the full generality of the problem considered in~\cite{Naor_etal_NCGI}, nevertheless $\alpha_{\CC}(d)^2$ is larger than this for all dimensions $d$.

Note also that to approximate (\ref{eq_QFOG}) with this approach one needs to have $N=dn$ in (\ref{NCGI_sdp}). This means that a na\"ive implementation of this relaxation would result in a semidefinite program with a matrix variable of size $d^2n^2\times d^2n^2$, while our approach is based on semidefinite programs with matrix variables of size $dn\times dn$. It is however conceivable that when restricted to problems of the type of (\ref{eq_QFOG}), the SDP relaxation (\ref{NCGI_sdp}) may enjoy certain symmetries or other properties that facilitate its solution.



\section{Applications}\label{section:applications}

Problem (\ref{eq_QFOG}) can describe several problems of interest. As examples, we describe below how it encodes a complementary version of the orthogonal Procrustes problem and the problem of Global Registration over Euclidean Transforms. Later, in Section \ref{Section:HigherRank}, we briefly discuss yet another problem, the Common Lines problem, that is encoded by a more general rank constrained version of (\ref{eq_QFOG}).

\subsection{Orthogonal Procrustes}

Given $n$ point clouds in $\RR^d$ of $k$ points each, the orthogonal Procrustes problem~\cite{Schonenmann66_Procrustes} consists of finding $n$ orthogonal transformations that best simultaneously align the point clouds. If the points are represented as the columns of matrices $A_1,\dots,A_n$, where $A_i \in \RR^{d\times k}$ then the orthogonal Procrustes problem consists of solving
\begin{equation}\label{eq:Procrustesmin}
\min_{O_1,.. .,O_n \in \OOO_d} \sum_{i,j=1}^n ||O_i^T A_i - O_j^T A_j||_F^2.
\end{equation}
Since $||O_i^T A_i - O_j^T A_j||_F^2 = \|A_i\|_F^2+\|A_j\|_F^2 - 2\tr\left( (A_iA_j^T)^T O_i O_j^T \right)$, (\ref{eq:Procrustesmin}) has the same solution as the complementary version of the problem
\begin{equation}\label{eq:Procrustesmaxdiag}
\max_{O_1,.. .,O_n \in \OOO_d} \sum_{i,j=1}^n \tr\left( (A_iA_j^T)^T O_i O_j^T \right).
\end{equation}
Since $C\in\RR^{dn\times dn}$ given by $C_{ij} = A_iA_j^T$ is positive semidefinite, problem (\ref{eq:Procrustesmaxdiag}) is encoded by (\ref{eq_QFOG}) and Algorithm~\ref{algorithm:main} provides a solution with an approximation ratio guaranteed (Theorem~\ref{lemma_maintheorem}) to be at least $\alpha_{\RR}(d)^2$.

The algorithm proposed in Naor et al.~\cite{Naor_etal_NCGI} gives an approximation ratio of $\frac1{2\sqrt{2}}$, smaller than $\alpha_{\RR}(d)^2$, for (\ref{eq:Procrustesmaxdiag}). As discussed above, the approach in \cite{Naor_etal_NCGI} is based on a semidefinite relaxation with a matrix variable of size $d^2n^2\times d^2n^2$ instead of $dn\times dn$ as in (\ref{eq_OCSDPv}) (see Section~\ref{sec:CompareNRV} for more details).

Nemirovski~\cite{Nemirovski_sumsProcrustes} proposed a different semidefinite relaxation (with a matrix variable of size $d^2n\times d^2n$ instead of $dn\times dn$ as in (\ref{eq_OCSDPv})) for the orthogonal Procrustes problem. In fact, his algorithm approximates the slightly different problem
\begin{equation}\label{eq:ProcrustesmaxNOdiag}
\max_{O_1,.. .,O_n \in \OOO_d} \sum_{i\neq j} \tr\left( (A_iA_j^T)^T O_i O_j^T \right),
\end{equation}
which is an additive constant (independent of $O_1,\dots, O_n$) smaller than (\ref{eq:Procrustesmaxdiag}). The best known approximation ratio for this semidefinite relaxation, due to So~\cite{So_sumsProcrustes},
 is $\OOO\left( \frac1{\log (n + k + d)} \right)$. Although an approximation to (\ref{eq:ProcrustesmaxNOdiag}) would technically be stronger than an approximation to (\ref{eq:Procrustesmaxdiag}), the two quantities are essentially the same provided that the point clouds are indeed perturbations of orthogonal transformations of the same original point cloud, as is the case in most applications (see~\cite{Naor_etal_NCGI} for a more thorough discussion on the differences between formulations (\ref{eq:Procrustesmaxdiag}) and (\ref{eq:ProcrustesmaxNOdiag})).

Another important instance of this problem is when the transformations are elements of $S\OOO_2$ (the special orthogonal group of dimension $2$, corresponding to rotations of the plane). Since $S\OOO_2$ is isomorphic to $\UUU_1$ we can encode it as an instance of problem (\ref{eq_QFUG}), in this case we recover the previously known optimal approximation ratio of $\frac{\pi}4$~\cite{So_Zhang_Ye_QuadSDPrelax}.

Note that, since all instances of problem (\ref{eq_QFOG}) can be written as an instance of orthogonal Procrustes, the integrality gap we show (Theorem~\ref{theoremtightnessratios}) guarantees that our approximation ratio is optimal for the natural semidefinite relaxation we consider for the problem.

\subsection{Global Registration over Euclidean Transforms}

The problem of global registration over Euclidean rigid motions is an extension of orthogonal Procrustes. In global registration, one is required to estimate the positions $x_1,\ldots,x_k$ of $k$ points in $\mathbb{R}^d$ and the unknown rigid transforms of $n$ local coordinate systems given (perhaps noisy) measurements of the local coordinates of each point in some (though not necessarily all) of the local coordinate systems. The problem differs from orthogonal Procrustes in two aspects: First, for each local coordinate system, we need to estimate not only an orthogonal transformation but also a translation in $\mathbb{R}^d$. Second, each point may appear in only a subset of the coordinate systems. Despite those differences, it is shown in~\cite{Chaudhury_etal_GlobalRegistration} that global registration can also be reduced to the form (\ref{eq_QFOG}) with a matrix $C$ that is positive semidefinite.

More precisely, denoting by $P_i$ the subset of points that belong to the $i$-th local coordinate system ($i = 1\dots n$), and given the local coordinates
$$x^{(i)}_l = O_i^T\left( x_l - t_i \right) + \xi_{il}$$
of point $x_l \in P_i$ (where $O_i$ denotes an unknown orthogonal transformation, $t_i$ an unknown translation and $\xi_{il}$ a noise term). The goal is to estimate the global coordinates $x_l$. The idea is to minimize the function
\begin{equation*}
\phi = \sum_{i=1}^n \sum_{l \in P_i} \left\| x_l - (O_i x_l^{(i)} + t_i) \right\|^2,
\end{equation*}
over $x_l,t_i\in \mathbb{R}^d,O_i \in \OOO_d$.
It is not difficult to see that the optimal $x_l^\star$ and $t_i^\star$ can be written in terms of $O_1,\dots,O_n$. Substituting them back into $\phi$, the authors in \cite{Chaudhury_etal_GlobalRegistration} reduce the previous optimization to solving
\begin{equation}
\label{opt324342}
\underset{O_i\in \OOO_d} \max \sum_{i=1}^{n}\sum_{j=1}^{n} \tr\left( \left[B L^\dagger B^T\right]_{ij} O_iO_j^T\right),
\end{equation}
where $L$ is a certain $(n+k)\times (n+k)$ Laplacian matrix, $L^\dagger$ is its pseudo inverse, and $B$ is a $(dn)\times (n+k)$ matrix (see~\cite{Chaudhury_etal_GlobalRegistration}). This means that $B L^\dagger B^T\succeq 0$, and (\ref{opt324342}) is of the form of~(\ref{eq_QFOG}).

\section{Analysis of the approximation algorithm}\label{Section:analysisalgorithm}

In this Section we prove Theorem \ref{lemma_maintheorem}. As~(\ref{eq_OCSDPv}) and~(\ref{eq_UCSDPv}) are relaxations of, respectively, problem~(\ref{eq_QFOG}) and problem~(\ref{eq_QFUG}) their maximums are necessarily at least as large as the ones of, respectively,~(\ref{eq_QFOG}) and~(\ref{eq_QFUG}). This means that Theorem \ref{lemma_maintheorem} is a direct consequence of the following Theorem.

\begin{theorem}\label{lemma_mainlemma}
Let $C\succeq 0 $ and real. Let $X_1,\dots,X_n$ be a feasible solution to (\ref{eq_OCSDPv}). Let $V_1,\dots,V_n\in \OOO_d$ be the output of the (random) rounding procedure described in Algorithm \ref{algorithm:main}. Then
\[
\EE \left[ \sum_{i=1}^n\sum_{j=1}^n\tr\left(C_{ij}^TV_iV_j^T\right) \right] \geq \alpha_{\RR}(d)^2 \sum_{i=1}^n\sum_{j=1}^n\tr\left(C_{ij}^TX_iX_j^T\right),
\]
where $\alpha_{\RR}(d)$ is the constant in Definition \ref{def:alphad}. Analogously, if $C\succeq 0 $ and complex and $Y_1,\dots,Y_n$ is a feasible solution of (\ref{eq_UCSDPv}) and $W_1,\dots,W_n\in \UUU_d$ the output of the (random) rounding procedure described in Algorithm \ref{algorithm:main}. Then
\[
\EE \left[ \sum_{i=1}^n\sum_{j=1}^n\tr\left(C_{ij}^HW_iW_j^H\right) \right] \geq \alpha_{\CC}(d)^2 \sum_{i=1}^n\sum_{j=1}^n\tr\left(C_{ij}^TY_iY_j^H\right),
\]
where $\alpha_{\CC}(d)$ is the constant in Definition \ref{def:alphad}.
\end{theorem}

In Section~\ref{section:integralitygap} we show that these ratios are optimal (Theorem~\ref{theoremtightnessratios}).

Before proving Theorem \ref{lemma_mainlemma} we present a sketch of the proof for the case $d=1$ (and real). The argument is known as the Rietz method (See~\cite{NAlon_ANaor_2006})~\footnote{These ideas also play a major role in the unidimensional complex case treated by So et al~\cite{So_Zhang_Ye_QuadSDPrelax}.}:

Let $X_1,\dots,X_n\in\RR^{1\times n}$ be a feasible solution to (\ref{eq_OCSDPv}), meaning that $X_iX_i^T = 1$. Let $R\in\RR^{n\times 1}$ be a random matrix with i.i.d. standard Gaussian entries. Our objective is to compare $\EE \left[\sum_{i,j}^nC_{ij}\sign(X_iR)\sign(X_jR)\right]$ with $\sum_{i,j}^nC_{ij}X_iX_j^T$. The main observation is that although $\EE \left[\sign(X_iR)\sign(X_jR)\right]$ is not a linear function of $X_iX_j^T$, the expectation $\EE\left[ \sign(X_iR)X_jR\right]$ is. In fact $\EE\left[ \sign(X_iR)X_jR\right]=\alpha_{\RR}(1)X_iX_j^T=\sqrt{\frac2{\pi}}X_iX_j^T$ --- which follows readily by thinking of $X_i$ and $X_j$ as vectors in the two dimensional plane that they span.  We use this fact (together with the positiveness of $C$) to show our result. The idea is to build the matrix $S\succeq 0$,
\[
S_{ij} = \left(X_iR - \sqrt{\frac{\pi}2}\sign(X_iR)\right)\left(X_jR - \sqrt{\frac{\pi}2}\sign(X_jR)\right).
\]
Since both $C$ and $S$ are PSD, $\tr(CS) \geq 0$, which means that
\[
0\leq \EE\left[ \sum_{ij}C_{ij} (X_iR - \sqrt{\frac{\pi}2}\sign(X_iR))(X_jR - \sqrt{\frac{\pi}2}\sign(X_jR))\right].
\]
Combining this with the observation above and the fact that $\EE\left[ X_iRX_jR\right] = X_iX_j^T$, we have
\[
\EE \sum_{i,j}^nC_{ij}\sign(X_iR)\sign(X_jR) \geq \frac2\pi \sum_{i,j}^nC_{ij}X_iX_j^T.
\]

\proof{[of Theorem~\ref{lemma_mainlemma}]
For the sake of brevity we restrict the presentation of the proof to the real case. Nevertheless, it is easy to see that all the arguments trivially adapt to the complex case by, essentially, replacing all transposes with Hermitian adjoints and $\alpha_{\RR}(d)$ with $\alpha_{\CC}(d)$.

Let $R\in\RR^{nd\times d}$ be a Gaussian random matrix with i.i.d entries $\NNN\left(0,\frac1d\right)$.
We want to provide a lower bound for
\[
\EE \left[ \sum_{i=1}^n\sum_{j=1}^n\tr\left(C_{ij}^TV_iV_j^T\right) \right] = \EE \left[ \sum_{i=1}^n\sum_{j=1}^n \tr\left( C_{ij}^T \PPP( U_i R ) \PPP( U_j R )^T \right)  \right].
\]
Similarly to the $d=1$ case, one of the main ingredients of the proof is the fact given by the lemma below.

\begin{lemma}\label{lemma_polar1_dequalr}
Let $r\geq d$. Let $M,N\in \RR^{d\times nd}$ such that $MM^T = NN^T = I_{d \times d}$. Let $R \in \RR^{nd \times d}$ be a Gaussian random matrix with real valued i.i.d entries $\NNN\left( 0, \frac1d \right)$. Then
\[
 \EE\left[ \PPP(M R) (N R)^T\right] = \EE\left[ (M R) \PPP(N R)^T\right] = \alpha_{\RR}(d)  MN^T,
\]
where $\alpha_{\RR}(d)$ is constant in Definition \ref{def:alphad}.

Analogously, if  $M,N\in \CC^{d\times nd}$ such that $MM^H = NN^H = I_{d \times d}$, and $R \in \CC^{nd \times r}$ is a Gaussian random matrix with complex valued i.i.d entries $\NNN\left( 0, \frac1d \right)$, then
\[ \EE\left[ \PPP(M R) (N R)^H\right] = \EE\left[ (M R) \PPP(N R)^H\right]  =\alpha_{\CC}(d)   MN^H,\]
where $\alpha_{\CC}(d)$ is constant in Definition \ref{def:alphad}.
\end{lemma}

Before proving Lemma~\ref{lemma_polar1_dequalr} we use it to finish the proof of Theorem~\ref{lemma_mainlemma}.

Just as above, we define the positive semidefinite matrix $S\in\RR^{dn \times dn}$ whose $(i,j)$-th block is given by
\[
S_{ij} = \left( U_i R - \alpha_{\RR}(d)^{-1} \PPP(U_iR)\right) \left( U_j R - \alpha_{\RR}(d)^{-1} \PPP(U_jR)\right)^T.
\]
We have $\EE S_{ij} =$ 
\begin{eqnarray*}
 & = & \EE\left[ U_iR  (U_jR)^T -  \alpha_{\RR}(d)^{-1} \PPP(U_iR)  (U_jR)^T  - \alpha_{\RR}(d)^{-1} U_iR  \PPP(U_jR)^T+ \alpha_{\RR}(d)^{-2}\PPP(U_iR)  \PPP(U_jR)^T \right] \\
     & = & U_i \EE\left[R R^T\right] U_j^T  -\alpha_{\RR}(d)^{-1} \EE\left[\PPP(U_iR)  (U_jR)^T\right] -  \alpha_{\RR}(d)^{-1} \EE\left[U_iR \PPP(U_jR)^T\right] + \alpha_{\RR}(d)^{-2}\EE\left[ V_i V_j^T\right] \\
 & = & U_i U_j^T -  U_i U_j^T - U_i U_j^T  + \alpha_{\RR}(d)^{-2}\EE\left[ V_i V_j^T\right] \\
 & = & \alpha_{\RR}(d)^{-2}\EE\left[ V_i V_j^T\right] -   U_i U_j^T.
\end{eqnarray*}

By construction $S\succeq 0$. Since $C\succeq 0 $, $\tr(CS)\geq 0$, which means that
\[
0\leq \EE\left[\tr \left(CS\right) \right]= \tr\left(C\EE[ S]\right)=\sum_{i=1}^n\sum_{j=1}^n\tr\left(  C_{ij}^T \left(\alpha_{\RR}(d)^{-2}\EE\left[ V_i V_j^T\right] -   U_i U_j^T\right) \right).
\]
Thus,
\[
\EE\left[\sum_{i=1}^n\sum_{j=1}^n\tr\left(  C_{ij}^TV_i V_j^T \right)\right] \geq \alpha_{\RR}(d)^{2} \sum_{i=1}^n\sum_{j=1}^n\tr\left(  C_{ij}^T U_i U_j^T\right).
\]
$\qed$}

We now present and prove an auxiliary lemma, needed for the proof of Lemma~\ref{lemma_polar1_dequalr}.

\begin{lemma}\label{lemma:alphad_dequalr}
Let $G$ be a $d\times d$ Gaussian random matrix with real valued i.i.d. $\NNN\left(0,\frac1d\right)$ entries and let $\alpha_{\RR}(d)$ as defined in Definition \ref{def:alphad}. Then, 
\[
\EE \left(\PPP(G) G^T\right)= \EE \left(G \PPP(G)^T\right)= \alpha_{\RR}(d)I_{d\times d}.
\]
Furthermore, if $G$ is a $d\times d$ Gaussian random matrix with complex valued i.i.d. $\NNN\left(0,\frac1d\right)$ entries and $\alpha_{\CC}(d)$ the analogous constant (Definition~\ref{def:alphad}), then
\[
\EE \left(\PPP(G) G^H\right)= \EE \left( G\PPP(G)^H\right)= \alpha_{\CC}(d)I_{d\times d}.\]
\end{lemma}

\proof{
We restrict the presentation to the real case.  All the arguments are equivalent to the complex case, replacing all transposes with Hermitian adjoints and $\alpha_{\RR}(d)$ with $\alpha_{\CC}(d)$.

Let $G = U \Sigma V^T$ be the singular value decomposition of $G$. Since $GG^T = U\Sigma^2 U^T$ is a Wishart matrix, it is well known that its eigenvalues and eigenvectors are independent and $U$ is distributed according to the Haar measure in $\OOO_d$ (see e.g. Lemma 2.6 in~\cite{Tulino_Verdu_RM}). To resolve ambiguities, we consider $\Sigma$ ordered such that $\Sigma_{11} \geq \Sigma_{22} \geq.. .\geq\Sigma_{dd}$.

Let $Y = \PPP(G) G^T$. Since
\[
\PPP(G) = \PPP(U  \Sigma V^T) = U I_{d\times d} V^T,
\]
we have
\[
Y = \PPP(U \Sigma V^T) (U \Sigma  V^T)^T = U I_{d\times d}V^T V \Sigma U^T = U \Sigma U^T.
\]
Note that $ G\PPP(G)^T = U \Sigma U^T =  Y$.

Denoting $u_1,\dots,u_d$ the rows of $U$, since $U$ is distributed according to the Haar measure, we have that $u_j$ and $-u_j$ have the same distribution conditioned on $\Sigma$ and $u_i$, for any $i\neq j$. This implies that if $i\neq j$, $Y_{ij} = u_i \Sigma u_j^T$ is a symmetric random variable, and so $\EE Y_{ij} = 0$. Also, $u_i\sim u_j$ implies that $Y_{ii}\sim Y_{jj}$. This means that $\EE Y = c I_{d\times d}$ for some constant $c$. To obtain $c$,
\[
c = c\frac1d\tr(I_{d\times d}) = \frac1d\EE \tr(Y) = \frac1d\EE \tr(U  \Sigma U^T) = \frac1d\EE\sum_{k=1}^n\sigma_{k}(G) = \alpha_{\RR}(d),
\]
which shows the lemma.
$\qed$}

\proof{[of Lemma~\ref{lemma_polar1_dequalr}]
We restrict the presentation of proof to the real case.  Nevertheless, as before, all the arguments trivially adapt to the complex case by, essentially, replacing all transposes with Hermitian adjoints and $\alpha_{\RR}(d)$ with $\alpha_{\CC}(d)$.

Let $A = \left[ M^T\text{  } N^T\right]\in\RR^{dn\times 2d}$ and $A=QB$ be the $QR$ decomposition of $A$ with  $Q\in \RR^{nd\times nd}$ an orthogonal matrix and $B \in \RR^{nd \times 2d}$ upper triangular with non-negative diagonal entries; note that only the first $2d$ rows of $B$ are nonzero. We can write
\[
Q^TA = B = \begin{bmatrix} B_{11} & B_{12}\\
0_d & B_{22}\\
0_d & 0_d\\\
\vdots & \vdots\\
0_d & 0_d
\end{bmatrix} \in \RR^{dn \times 2d},
\]
where $B_{11}\in\RR^{d\times d}$ and $B_{22}\in\RR^{d\times d}$ are upper triangular matrices with non-negative diagonal entries. Since $$(Q^T M^T)^T_{11}(Q^T M^T)_{11} = (Q^T M^T)^T(Q^T M^T) = M Q Q^T M^T = M I_{nd \times nd} M^T = M M^T = I_{d \times d},$$ $B_{11} = (Q^T M^T)_{11}$ is an orthogonal matrix, which together with the non-negativity of the diagonal entries (and the fact that $B_{11}$ is upper-triangular) forces $B_{11}$ to be $B_{11} = I_{d\times d}$.

Since $R$ is a Gaussian matrix and $Q$ is an orthogonal matrix, $QR \sim R$ which implies
\[
 \EE\left[ \PPP(M R) (N R)^T\right] = \EE \left[ \PPP(M Q R)(NQ R)^T \right].
\]
Since $MQ=[B_{11}^T, 0_{d\times d},\cdots,0_{d\times d}] = [I_{d\times d}, 0_{d\times d},\cdots,0_{d\times d}]$ and $NQ = [B_{12}^T, B_{22}^T, 0_{d\times d},\cdots,0_{d\times d}]$,
\[
\EE\left[ \PPP(M R) (N R)^T\right] =  \e \left[\PPP(R_1)(B_{12}^T R_1 + B_{22}^T R_2)^T\right],
\]
where $R_1$ and $R_2$ are the first two $d\times d$ blocks of $R$. Since these blocks are independent, the second term vanishes and we have
\[
\EE\left[ \PPP(M R) (N R)^T\right] =  \EE\left[ \PPP(R_1) R_1^T\right]B_{12}.
\]
The Lemma now follows from using Lemma \ref{lemma:alphad_dequalr} to obtain $\EE \left[\PPP(R_1) R_1^ T\right]= \alpha_{\RR}(d)I_{d\times d}$ and noting that $B_{12} = (Q^TM^T)^T(Q^TN^T) = MN^T$.

The same argument, with $Q'B'$ the $QR$ decomposition of $A' = \left[ N^T M^T\right]\in\RR^{dn\times 2d}$ instead, shows
\[
\EE\left[ (M R) \PPP(N R)^T\right] = \EE \left[  R_1\PPP(R_1)^T\right] MN^T = \alpha_{\RR}(d)MN^T.
\]

$\qed$}


\section{The approximation ratios $\alpha_{\RR}(d)^2$ and $\alpha_{\CC}(d)^2$}\label{section:onalphad}


The approximation ratio we obtain (Theorem~\ref{lemma_maintheorem}) for Algorithm \ref{algorithm:main} is given, in the orthogonal case, by $\alpha_{\RR}(d)^2$ and, in the unitary case, by $\alpha_{\CC}(d)^2$. $\alpha_{\RR}(d)$ and $\alpha_{\CC}(d)$ are defined as the average singular value of a $d\times d$ Gaussian matrix $G$ with, respectively real and complex valued, i.i.d $\NNN(0,\frac1d)$ entries. These singular values correspond  to the square root of the eigenvalues of a Wishart matrix $W = G G^T$, which are well-studied objects (see, e.g., \cite{Shen01_Wishart} or \cite{Couillet_RMbook}).


For $d=1$, this corresponds to the expected value of the absolute value of standard Gaussian (real or complex) random variable. Hence,
\[
\alpha_{\RR}(1) =  \sqrt{\frac2\pi} \text{ and } \alpha_{\CC}(1) = \sqrt{\frac{\pi}4},
\]
meaning that, for $d=1$, we recover the approximation ratio of $\frac2\pi$,  of Nesterov~\cite{Nesterov_quadprogram1} for the real case, and the approximation ratio of $\frac{\pi}4$ of So et al.~\cite{So_Zhang_Ye_QuadSDPrelax} in the complex case.

For any $d\geq 1$, the marginal distribution of an eigenvalue of the Wishart matrix $W = G G^T$ is known~\cite{Livan11_WishartLaguerre,Couillet_RMbook,Oliver_lecturenotesRM} (see Section~\ref{sec:appendixbounds}). Denoting by $p_d^{(\KK)}$ the marginal distribution for $\KK = \RR$ and $\KK = \CC$, we have
\begin{equation}\label{integral:alphak}
\alpha_{\KK}(d) = \frac{1}{d^{1/2}} \int_0^{\infty} x^{1/2} p_d^{(\KK)}(x) dx.
\end{equation}
In the complex valued case, $p_d^{(\CC)}(x)$ can be written in terms of Laguerre polynomials~\cite{Couillet_RMbook,Oliver_lecturenotesRM} and $\alpha_{\CC}(d)$ is given by
\begin{equation} \label{sum:alphaC}
\alpha_{\CC}(d) = d^{-3/2}\sum_{n=0}^{d-1}\int_{0}^\infty x^{1/2}e^{-x}L_n(x)^2 dx,
\end{equation}
Where $L_n(x)$ is the $n$th Laguerre polynomial.
In Section~\ref{sec:appendixbounds} we give a lower bound to (\ref{sum:alphaC}). The real case is more involved~\cite{Livan11_WishartLaguerre}, nevertheless we are able to provide a lower bound for $\alpha_{\RR}(d)$ as well.
\begin{theorem}\label{lemma:alphadbounds}
Consider $\alpha_{\RR}(d)$ and $\alpha_{\CC}(d)$ as defined in (\ref{def:alphad}).  The following holds,
\[
\alpha_{\CC}(d) \geq \frac8{3\pi} - \frac{5.05}d \text{ and } \alpha_{\RR}(d) \geq \frac8{3\pi}  - \frac{9.07}d.
\]
\end{theorem}
\proof{
These bounds are a direct consequence of Lemmas \ref{lemma:alphacbounds} and \ref{lemma:alpharcdiff}.
}

%
%

One can easily evaluate $\lim_{d\to\infty}\alpha_{\KK}(d)$ (without using Theorem~\ref{lemma:alphadbounds}) by noting that the distribution of the eigenvalues of the Wishart matrix we are interested in, as $d\to\infty$, converges in probability to the Marchenko-Pastur distribution~\cite{Shen01_Wishart} with density
\[
\mathrm{mp}(x) = \frac{1}{2\pi x} \sqrt{x(4-x)} \mathbf{1}_{[0,4]},
\]
for both $\KK =\RR$ and $\KK = \CC$. This immediately gives,
\[
\lim_{d\to\infty}\alpha_{\KK}(d) = \int_0^4 \sqrt{x} \frac{1}{2\pi x} \sqrt{x(4-x)}  dx = \frac{8}{3\pi}.
\]
We note that one could also obtain lower bounds for $\alpha_{\KK}^2(d)$ from results on the rate of convergence to $\mathrm{mp}(x)$~\cite{Gotze_ConvergenceMP}. However this approach seems to not provide bounds with explicit constants and to not be as sharp as the approach taken in Theorem~\ref{lemma:alphadbounds}.

\begin{figure}[h]
\centering
\includegraphics[width=1\textwidth]{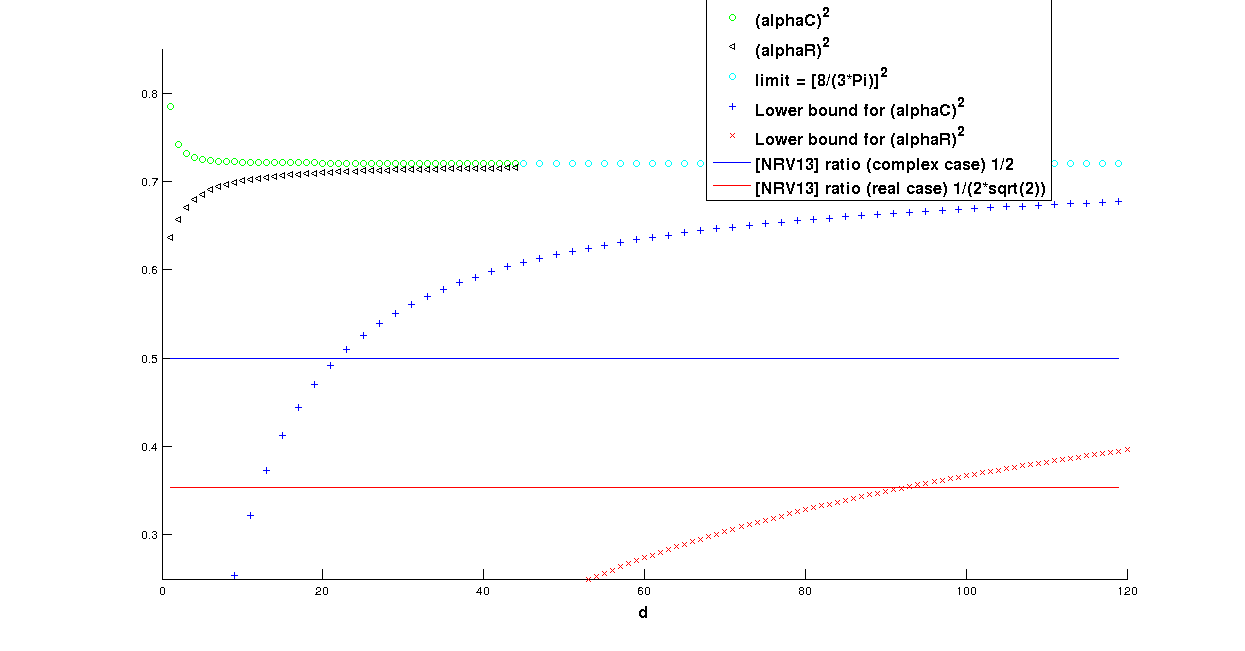}
\caption{Plot showing the computed values of $\alpha_{\KK}(d)^2$, for $d\leq 44$, the limit of $\alpha_{\KK}(d)^2$ as $d\to\infty$, the lower bound for $\alpha_{\KK}(d)^2$ given by Theorem~\ref{lemma:alphadbounds} as function of $d$, and the approximation ratio of $\frac1{2\sqrt{2}}$ and $\frac12$ obtained in~\cite{Naor_etal_NCGI}.}
\label{figure:alphad}
\end{figure}

For any $d$, the exact value of $\alpha_{\KK}(d)$ can be computed, by (\ref{integral:alphak}), using {\tt Mathematica} (See table below). Figure~\ref{figure:alphad} plots these values for $d=1,\dots,44$. We also plot the bounds for the real and complex case obtained in Theorem~\ref{lemma:alphadbounds}, and the approximation ratios obtained in~\cite{Naor_etal_NCGI}, for comparison.

\begin{center}

\begin{tabular}{ccccccc}\label{table:alphad}

$d$ & $\alpha_{\RR}(d)$ & $\alpha_{\CC}(d)$ & $\alpha_{\RR}(d) \approx$& $\alpha_{\RR}(d)^2\approx$ & $\alpha_{\CC}(d) \approx$ & $\alpha_{\CC}(d)^2 \approx$ \\

\hline

$1$ &  $\sqrt{\frac2\pi}$ & $\frac{\sqrt{\pi}}2$ & $0.7979$ & 0.6366 &0.8862 &0.7854\\

$2$ &  $\frac{2\sqrt{2}-1}4\sqrt{\pi} $ &$\frac{11\sqrt{\frac{\pi}2}}{16}$ & $0.8102$ & 0.6564 &0.8617 &0.7424 \\

$3$ &  $\frac{2\sqrt{2}+3\pi}{6\sqrt{3\pi}}$&$\frac{107 \sqrt{\frac{\pi}{3}}}{128}$ & $0.8188$ & 0.6704 &0.8554 &0.7312 \\

$\infty$ & $\frac8{3\pi}$ & $\frac8{3\pi}$ & $0.8488$ & 0.7205 & $0.8488$ & 0.7205

\end{tabular}

\end{center} 

The following conjecture is suggested by our analysis and numerical computations.
\begin{conjecture}\label{conjecture:alphar}
Let $\alpha_{\RR}(d)$ and $\alpha_{\CC}(d)$ be the average singular value of a $d\times d$ matrix with random i.i.d., respectively real valued and complex valued, $\NNN\left(0,\frac1d\right)$ entries (see Definition~\ref{def:alphad}). Then, for all $d\geq 1$,
\[
\alpha_{\CC}(d+1) \leq \alpha_{\CC}(d) \text{ and }\alpha_{\RR}(d+1) \geq \alpha_{\RR}(d),
\]
\end{conjecture}

\section{The little Grothendieck problem over the Stiefel manifold}\label{Section:HigherRank}

In this section we focus on a generalization of (\ref{eq_QFOG}), the little Grothendieck problem over the Stiefel manifold $\OOO_{(d,r)}$, the set of matrices $O\in\RR^{d\times r}$ such that $OO^T = I_{d\times d}$. In this exposition we will restrict ourselves to the real valued case but it is easy to see that the ideas in this Section easily adapt to the complex valued case.

We consider the problem
\begin{equation}
\max_{O_1,.. .,O_n \in \OOO_{(d,r)}} \sum_{i=1}^n\sum_{j=1}^n\tr\left(C_{ij}^TO_iO_j^T\right),
\label{eq_QFOGr}
\end{equation}
for $C\succeq0$. The special case $d=1$ was formulated and studied in~\cite{Briet09generalizedGrothendieckXOR} and~\cite{Jop_littleGTrank} in the context of quantum non-locality and quantum XOR games.

Note that, for $r=d$, problem (\ref{eq_QFOGr}) reduces to (\ref{eq_QFOG}) and, for $r=nd$, it reduces to the tractable relaxation (\ref{eq_OCSDPv}). As a solution to (\ref{eq_QFOG}) can be transformed, via zero padding, into a solution to (\ref{eq_QFOGr}) with the same objective function value, Algorithm~\ref{algorithm:main} automatically provides an approximation ratio for (\ref{eq_QFOGr}), however we want to understand how this approximation ratio can be improved using the extra freedom (in particular, in the case $r=nd$, the approximation ratio is trivially $1$). Below we show an adaptation of Algorithm~\ref{algorithm:main}, based on the same relaxation (\ref{eq_OCSDPv}), for problem (\ref{eq_QFOGr}) and show an improved approximation ratio.

\begin{algorithm}\label{algorithm:rankr}
 Compute $X_1,\dots,X_n\in \RR^{d\times nd}$  a solution to (\ref{eq_OCSDPv}). Let $R$ be a $nd\times r$ Gaussian random matrix whose entries are real i.i.d. $\NNN(0,\frac1r)$.
The approximate solution for (\ref{eq_QFOGr}) is now computed as
\[
V_i = \PPP_{(d,r)}(X_i R),
\]
where $\PPP_{(d,r)}(X)=\argmin_{Z\in \OOO_{(d,r)} }\|Z-X\|_F$, for any $X\in\RR^{d\times r}$, is a generalization of the polar component to the Stiefel manifold $\OOO_{(d,r)}$.
\end{algorithm}

Below we show an approximation ratio for Algorithm~\ref{algorithm:rankr}.

\begin{theorem}\label{thm:alg:rankr}
Let $C\succeq 0$. Let $V_1,\dots,V_n\in \OOO_{(d,r)}$ be the (random) output of Algorithm~\ref{algorithm:rankr}. Then,
\[
\EE \left[ \sum_{i=1}^n\sum_{j=1}^n\tr\left(C_{ij}^TV_iV_j^T\right) \right] \geq \alpha_{\RR}(d,r)^2 \max_{O_1,.. .,O_n \in \OOO_{(d,r)}} \sum_{i=1}^n\sum_{j=1}^n\tr\left(C_{ij}^TO_iO_j^T\right),
\]
where $\alpha_{\RR}(d,r)$ is the defined below (Definition~\ref{def:alphadr}).
\end{theorem}

\begin{definition}\label{def:alphadr}
Let $r\geq d$ and $G \in \RR^{d\times r}$ be a Gaussian random matrix with i.i.d real entries $\NNN\left(0,\frac1r\right)$. We define
\[
\alpha_{\RR}(d,r) := \e\left[ \frac1d \sum_{j=1}^d \sigma_j(G)\right],
\]
where $\sigma_j(G)$ is the $j$th singular value of $G$.
\end{definition}
We investigate the limiting behavior of $\alpha_{\RR}(d,r)$ as $r\to \infty$ and as $r,d\to \infty$ at a proporitional rate in Section 6.2.

For the sake of brevity we omit the proof of this Theorem. We do state and prove Lemmas~\ref{lemma:alphad} and~\ref{lemma_polar1} on the Appendix, which are the analogous, to this setting, of Lemmas~\ref{lemma:alphad_dequalr} and~\ref{lemma_polar1_dequalr}. It is then not difficult to see that the arguments in the proof of Theorem~\ref{lemma_mainlemma} trivially adapt to this case and that the proof of Theorem~\ref{thm:alg:rankr} is completely analogous to the one of Theorem~\ref{lemma_maintheorem}.


Besides the applications, for $d=1$, described in~\cite{Briet09generalizedGrothendieckXOR} and~\cite{Jop_littleGTrank}, Problem (\ref{eq_QFOGr}) is also motivated by an application in molecule imaging, the common lines problem.

\subsection{The common lines problem}

The common lines problem arises in three-dimensional structure determination of biological molecules using Cryo-Electron Microscopy~\cite{ASinger_YShkolnisky_commonlines}, and can be formulated as follows.
Consider $n$ rotation matrices $O_1,\ldots,O_n \in S\OOO_3$. The
three columns of each rotation matrix form a orthonormal basis
to $\mathbb{R}^3$. In particular, the first two columns of each rotation
matrix span a two-dimensional subspace (a plane) in $\mathbb{R}^3$. We
assume that no two planes are parallel. Every pair of planes intersect at a
line, called the common-line of intersection. Let $b_{ij}\in \mathbb{R}^3$
be a unit vector that points in the direction of the common-line between the
planes corresponding to $O_i$ and $O_j$. Hence, there exist unit vectors
$c_{ij}$ and $c_{ji}$ with vanishing third component (i.e., $c_{ij} =
(x_{ij}, y_{ij}, 0)^T$) such that $O_i c_{ij} = O_j c_{ji} = b_{ij}$. The
common lines problem consists of estimating the rotation matrices
$O_1,\ldots,O_n$ from (perhaps noisy) measurements of the unit vectors
$c_{ij}$ and $c_{ji}$. The least-squares formulation of this problem is
equivalent to
\begin{equation}\label{CL_eq1}
\max_{O_1,\ldots,O_n \in S\OOO_3} \sum_{i,j=1}^n \tr(c_{ji}c_{ij}^T O_i^TO_j)
\end{equation}
However, since $c_{ij}$ has zero in the third coordinate, the common-line
equations  $O_i c_{ij} = O_j c_{ji}$ do not involve the third columns of the
rotation matrices.  The optimization problem (\ref{CL_eq1}) is therefore equivalent to
\begin{equation}\label{CL_eq2}
\max_{\tilde{O}_1^T,\ldots,\tilde{O}_n^T \in \OOO_{(2,3)}} \sum_{i,j=1}^n\tr(\Pi(c_{ji})\Pi(c_{ij})^T \tilde{O}_i^T \tilde{O}_j),
\end{equation}
where $\Pi :\mathbb{R}^3\to \mathbb{R}^2$ is a projection discarding the third component
(i.e., $\Pi(x,y,z) = (x,y)$) and $\tilde{O}_i^T\in \OOO_{(2,3)}$. 
The coefficient matrix in (\ref{CL_eq2}), $C_{ij} = \Pi(c_{ij})\Pi(c_{ji})^T$, is not positive semidefinite. However, one can add a diagonal matrix with large enough values to it in order to make it PSD. Although this does not affect the solution of (\ref{CL_eq2}) it does increase its function value by a constant, meaning that the approximation ratio obtained in Theorem~\ref{thm:alg:rankr} does not directly translate into an approximation ratio for Problem (\ref{CL_eq2}); see Section~\ref{section:futurework} for a discussion on extending the results to the non positive semidefinite case.


%

\subsection{The approximation ratio $\alpha_{\RR}(d,r)^2$}

In this Section we attempt to understand the behavior of $\alpha_{\RR}(d,r)^2$, the approximation ratio obtained for Algorithm~\ref{algorithm:rankr}. Recall that $\alpha_{\RR}(d,r)$ is defined as the average singular value of $G \in \RR^{d\times r}$, a Gaussian random matrix with i.i.d. entries $\NNN\left(0,\frac1r\right)$.

For $d=1$ this simply corresponds to the average length of a Gaussian vector in $\RR^r$ with i.i.d. entries $\NNN\left(0,\frac1r\right)$. This means that $\alpha_{\RR}(1,r)$ is the mean of a normalized chi-distribution,
$$\alpha_{\RR}(1,r) = \sqrt{\frac2r}\frac{\Gamma\left(\frac{r+1}2\right)}{\Gamma\left(\frac{r}2\right)}.$$
In fact, this corresponds to the results of Briet el al~\cite{Jop_littleGTrank}, which are known to be sharp~\cite{Jop_littleGTrank}.

For $d>1$ we do not completely understand the behavior of $\alpha_{\RR}(d,r)$, nevertheless it is easy to provide a lower bound for it by a function approaching $1$ as $r \to \infty$.
\begin{proposition}
Consider $\alpha_{\RR}(d,r)$ as in Definition~\ref{def:alphadr}. Then,
\begin{equation}\label{trivialsingularvaluebound}
\alpha_{\RR}(d,r) \geq 1 - \sqrt{\frac{d}r}.
\end{equation}
\end{proposition}
\proof{Gordon's theorem for Gaussian matrices (see Theorem 5.32 in ~\cite{VershyninNARandomMatrices})} gives us
\[
\EE s_{min}(G) \geq 1 - \sqrt{\frac{d}r},
\]
where $s_{min}(G)$ is the smallest singular value.  The bound follows immediately from noting that the average singular value is larger than the expected value of the smallest singular value.

As we are bounding $\alpha_{\RR}(d,r)$ by the expected value of the smallest singular value of a Gaussian matrix, we do not expect (\ref{trivialsingularvaluebound}) to be tight. In fact, for $d=1$, the stronger $\alpha_{\RR}(1,r) \geq 1 - \OOO\left(\frac{1}r\right)$ bound holds~\cite{Jop_littleGTrank}.

Similarly to $\alpha_{\RR}(d)$, we can describe the behavior of $\alpha_{\RR}(d,r)$ in the limit as $d\to \infty$ and $\frac{r}{d}\to \rho$. More precisely, the singular values of $G$ correspond to the square root of the eigenvalues of the Wishart matrix~\cite{Couillet_RMbook} $GG^T\sim W_d\left(\frac1{r},r\right)$. Let us set $r = \rho d$, for $\rho\geq 1$. The distribution of the eigenvalues of a Wishart matrix $W_d\left(\frac1{\rho d},d\right)$, as $d\to\infty$ are known to converge to the Marchenko Pastur distribution (see~\cite{Couillet_RMbook}) given by
\[
 d\nu(x) = \frac1{2\pi}\frac{\sqrt{((1+\sqrt{\lambda})^2-x)(x-(1-\sqrt{\lambda})^2)}}{\lambda x}\mathbf{1}_{[(1-\sqrt{\lambda})^2,(1+\sqrt{\lambda})^2]}dx,
\]
where $\lambda = \frac1{\rho}$.

Hence, we can define $\phi(\rho)$ as
\[
 \phi(\rho) := \lim_{d\to\infty}\alpha_{\RR}(d,\rho d)  =  \int_{\left(1-\sqrt{\frac1\rho}\right)^2}^{\left(1+\sqrt{\frac1\rho}\right)^2}\sqrt{x} \frac1{2\pi}\frac{\sqrt{(\left(1+\sqrt{\frac1\rho}\right)^2-x)(x-\left(1-\sqrt{\frac1\rho}\right)^2)}}{\frac1\rho x}dx.
\]
Although we do not provide a closed form solution for $\phi(\rho)$ the integral can be easily computed numerically and we plot it below. It shows how the approximation ratio improves as $\rho$ increases.

\begin{figure}[h]
\centering
\includegraphics[width=0.7\textwidth]{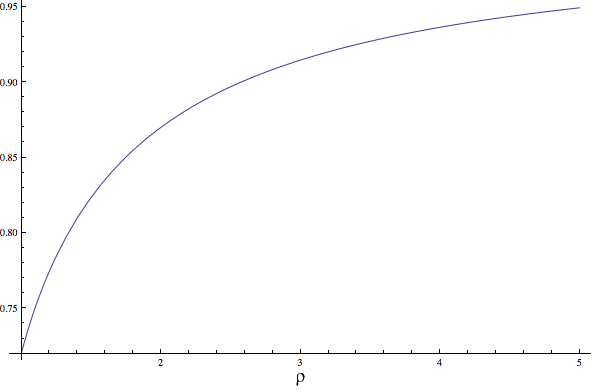}
\caption{Plot of $\phi(\rho) = \lim_{d\to\infty}\alpha_{\RR}(d,\rho d)$ for $\rho\in [1,5]$.}
\label{figure:psirho}
\end{figure}

\section{Integrality Gap}\label{section:integralitygap}

In this section we provide an integrality gap for relaxation (\ref{eq_OCSDPv}) that matches our approximation ratio $\alpha_{\RR}(d)^2$. For the sake of the exposition we will restrict ourselves to the real case, but it is not difficult to see that all the arguments can be adapted to the complex case.

Our construction is an adaption of the classical construction for the $d=1$ case (see, e.g., \cite{NAlon_ANaor_2006}). As it will become clear below, there is an extra difficulty in the $d>1$ orthogonal case. In fact, the bound on the integrality gap of (\ref{eq_OCSDPv}) given by this construction is $\alpha_{\RR}^\ast(d)^2$, defined as
\begin{equation}\label{def:alphadstarREAL}
 \alpha_{\RR}^\ast(d) = \max_{ \substack{D \text{ diagonal} \\ \|D\|_F^2 = d ,\ D \succeq 0 } }\EE \frac1d\sum_{i=1}^d\sigma_i(GD),
\end{equation}
where $G$ is a Gaussian matrix with i.i.d. real entries $\NNN\left(0,\frac1d\right)$.

Fortunately, using the notion of operator concavity of a function and the Lowner-Heinz Theorem~\cite{Carlen_QuantumEntropy}, we are able to show the following Theorem.

\begin{theorem}\label{alphastaralpha}
Let $d\geq 1$. Also, let $\alpha_{\RR}(d)$ be as defined in Definition~\ref{def:alphad} and $\alpha_{\RR}^\ast(d)$ as defined in~(\ref{def:alphadstarREAL}). Then,
\[
\alpha_{\RR}^\ast(d) = \alpha_{\RR}(d).
\]
\end{theorem}

\proof{
We want to show that 
\begin{equation*}
\max_{ \substack{D \text{ diagonal} \\ \|D\|_F^2 = d \\ D \succeq 0 } }\EE \sum_{i=1}\sigma_i(GD) = \EE \sum_{i=1}\sigma_i(G),
\end{equation*}
where $G$ is a $d\times d$ matrix with i.i.d. entries $\NNN\left(0,\frac1d\right)$. By taking $V = D^2$, and recalling the definition of singular value, we obtain the following claim (which immediately implies Theorem~\ref{alphastaralpha})
\begin{claim}\label{eq:appendixalphastarstuff}
\[\max_{ \substack{V \text{ diagonal} \\ \tr(V) = d \\ V \succeq 0 } }\EE \tr\left[\left( GVG^T \right)^{\frac12}\right]= \EE \tr\left[\left( GG^T \right)^{\frac12}\right].
\]
\end{claim}
\proof{
We will proceed by contradiction, suppose (\ref{eq:appendixalphastarstuff}) does not hold.  Since the optimization space is compact and the function continuous it must have a maximum that is attained by a certain $V\neq I_{d\times d}$. Out of all maximizers $V$, let $V^{(\ast)}$ be the one with smallest possible Frobenius norm.  The idea will be to use concavity arguments to build an optimal $V^{(\card)}$ with smaller Frobenius norm, arriving at a contradiction and hence showing the theorem.

Since $V^{(\ast)}$ is optimal we have
\[
\EE \tr\left[\left( GV^{(\ast)} G^T \right)^{\frac12}\right] = \alpha_{\RR}^\ast(d).
\]
Furthermore, since $V^{(\ast)}\neq I_{d\times d}$, it must have two different diagonal elements. Let $V^{(\ast\ast)}$ be a matrix obtained by swapping, in $V^{(\ast)}$, two of its non-equal diagonal elements. Clearly, $\|V^{(\ast\ast)}\|_F = \|V^{(\ast)}\|_F$ and, because of the rotation invariance of the Gaussian, it is easy to see that
\[
\EE \tr\left[\left( GV^{(\ast\ast)} G^T \right)^{\frac12}\right] = \alpha_{\RR}^\ast(d).
\]
Since $V^{(\ast)}\succeq 0$, these two matrices are not multiples of each other and so
\[
V^{(\card)} = \frac{V^{(\ast)}+V^{(\ast\ast)}}2,
\]
has a strictly smaller Frobenius norm than $V^{(\ast)}$. It is also clear that $V^{(\card)}$ is a feasible solution. We conclude the proof by showing
\begin{equation}\label{eq:laststepalphastart22}
\EE \tr\left[\left( GV^{(\card)} G^T \right)^{\frac12}\right] \geq \frac12\left( \EE \tr\left[\left( GV^{(\ast)} G^T \right)^{\frac12}\right] + \EE \tr\left[\left( GV^{(\ast\ast)} G^T \right)^{\frac12}\right] \right)= \alpha_{\RR}^\ast(d).
\end{equation}
By linearity of expectation and construction of $V^{(\card)}$, (\ref{eq:laststepalphastart22}) is equivalent to
\[
\EE\left[\tr\left[\left( \frac{GV^{(\ast)}G^T+GV^{(\ast\ast)}G^T }2 \right)^{\frac12}\right] - \frac12\left( \tr\left[\left( GV^{(\ast)} G^T \right)^{\frac12}\right] + \tr\left[\left( GV^{(\ast\ast)} G^T \right)^{\frac12}\right] \right) \right]\geq 0.
\]
This inequality follows from the stronger statement: Given two $d\times d$ matrices $A\succeq 0$ and $B\succeq 0$, the following holds 
\begin{equation}\label{concavityofsquareroot}
\left( \frac{A + B}2 \right)^{\frac12} - \frac{ A ^{\frac12} + B^{\frac12}  }2 \succeq 0.
\end{equation}
Finally, (\ref{concavityofsquareroot}) follows from the Lowner-Heinz Theorem, which states that the square root function is operator concave (See these lecture notes~\cite{Carlen_QuantumEntropy} for a very nice introduction to these inequalities).
}
}

Theorem~\ref{alphastaralpha} guarantees the optimality of the approximation ratio obtained in Section~\ref{Section:analysisalgorithm}. In fact, we show the theorem below.

\begin{theorem}\label{theoremtightnessratios}
For any $d\geq 1$ and any $\varepsilon >0$, there exists $n$ for which there exists $C\in\RR^{dn\times dn}$ such that $C\succeq 0$, and 
\begin{equation}\label{condition14}
\frac{\displaystyle{\max_{O_1,.. .,O_n \in \OOO_d} \sum_{i=1}^n\sum_{j=1}^n\tr\left(C_{ij}^TO_iO_j^T\right)}}{\displaystyle{\max_{\substack{ X_iX_i^T=I_{d\times d} \\ X_i\in\RR^{d\times dn} }} \sum_{i=1}^n\sum_{j=1}^n\tr\left(C_{ij}^TX_iX_j^T\right)}}\leq  \alpha_{\RR}(d)^2 + \varepsilon.
\end{equation}
\end{theorem}

We will construct $C$ randomly and show that it satisfies (\ref{condition14}) with positive probability. Given $p$ an integer we consider $n$ i.i.d. matrix random variables $V_k$, with $k=1,\dots,n$, where each $V_k$ is a $d\times dp$ Gaussian matrix whose entries are $\NNN(0,\frac1{dp})$. We then define $C$ as the random matrix with $d\times d$ blocks $C_{ij} = \frac1{n^2}V_iV_j^T$.
The idea now is to understand the typical behavior of both
\[
 w_r = \max_{\substack{ X_iX_i^T=I_{d\times d} \\ X_i\in\RR^{d\times dn} }} \sum_{i=1}^n\sum_{j=1}^n\tr\left(C_{ij}^TX_iX_j^T\right) \text{ and } w_c = \max_{O_1,.. .,O_n \in \OOO_d} \sum_{i=1}^n\sum_{j=1}^n\tr\left(C_{ij}^TO_iO_j^T\right).
\]

For $w_c$, we can rewrite
\[
w_c = \max_{O_1,.. .,O_n \in \OOO_d} \frac1{n^2}\sum_{i,j}\tr\left((V_iV_j^T)^TO_iO_j^T\right) = \max_{O_1,.. .,O_n \in \OOO_d}\left\|\frac1n\sum_{i=1}^n O_i^T V_i   \right\|_F^2.
\]
If $\DDD = \frac{\frac1n\sum_{i=1}^n O_i^T V_i }{\left\|\frac1n\sum_{i=1}^n O_i^T V_i \right\|_F  }$ then $\sqrt{w_c} = \sum_{i=1}^n \tr\left(\max_{O_i\in\OOO_d} O_i^TV_i\DDD^T \right) = \sum_{i=1}^n \tr\left( \PPP(V_i\DDD^T)^TV_i\DDD^T \right)$.
The idea is that, given a fixed (direction unit frobenius-norm matrix) $\DDD$, $\sum_{i=1}^n \tr\left( \PPP(V_i\DDD^T)^TV_i\DDD^T \right)$ converges to the expected value of one of the summands and, by an $\epsilon$-net argument (since the dimension of the space where $\DDD$ is depends only on $d$ and $p$ and the number of summands is $n$ which can be made much larger than $d$ and $p$) we can argue that the sum is close, for all $\DDD$'s simultaneously, to that expectation. It is not hard to see that we can assume that $\DDD = \frac1{\sqrt{d}}[D\ 0]$ where $D$ is diagonal and non-negative $d\times d$ matrix with $\|D\|_F^2 = d$. In that case (see (\ref{def:alphadstarREAL})),
\[ \EE \tr\left( \PPP(V_i\DDD^T)^TV_i\DDD^T \right) = \EE \frac1{\sqrt{pd}}\sum_{k=1}^d\sigma_k(GD) \leq \sqrt{\frac{d}{p}}\alpha_{\RR}^\ast(d), \]
 where $G$ is a Gaussian matrix with i.i.d. real entries $\NNN\left(0,\frac1d\right)$. This, together with Theorem~\ref{alphastaralpha}, gives \( \EE \tr\left( \PPP(V_i\DDD^T)^TV_i\DDD^T \right) \leq \sqrt{\frac{d}{p}}\alpha_{\RR}(d)  \).
All of this is made precise in the following lemma
\begin{lemma}\label{lemma:wcnotappendix}
For any $d$ and $\varepsilon>0$ there exists $p_0$ and $n_0$ such that, for any $p>p_0$ and $n>n_0$,
\[
 \max_{O_1,.. .,O_n \in \OOO_d}\left\|\frac1n\sum_{i=1}^n O_i^T V_i   \right\|_F^2 \leq \frac{d}p\alpha_{\RR}(d)^2 + \varepsilon,
\]
with probability strictly larger than $1/2$.
\end{lemma}

\proof{
Let us define
\[
A(V) :=  \max_{O_1,.. .,O_n \in \OOO_d}\left\|\frac1n\sum_{i=1}^n O_i^T V_i   \right\|_F.
\]
We have
\begin{eqnarray*}
A(V)  & = & \max_{D\in\RR^{d\times pd}: \|D\|_F = 1}\max_{O_1,.. .,O_n \in O(d)} \tr\left(\frac1n \sum_{i=1}^nO_i^TV_i D^T \right) \\ & = & \max_{D\in\RR^{d\times pd}: \|D\|_F = 1}  \frac1n\sum_{i=1}^n\max_{O_i\in O(d)}\tr\left(O_i^TV_i D^T \right) \\ & = &  \max_{D\in\RR^{d\times pd}: \|D\|_F = 1}  \frac1n\sum_{i=1}^n\tr\left(\PPP\left(V_i D^T \right)^TV_i D^T \right).
\end{eqnarray*}

For $D$ with $\|D\|_F=1$, we define
\[
A_{D}(V) = \frac1n\sum_{i=1}^n\tr\left(\PPP\left(V_i D^T \right)^TV_i D^T \right).
\]
We proceed by understanding the behavior of $A_{D}(V)$ for a specific $D$.

Let $D = U_L[\Sigma \ 0]U_R^T$, where $\Sigma$ is a $d\times d$ non-negative diagonal matrix, be the singular value decomposition of $D$. For each $i=1,\dots,n$, we have (using rotation invariance of the Gaussian distribution):
\begin{eqnarray*}
\tr\left(\PPP\left(V_iD^T \right)^TV_i D^T \right) & \sim & \tr\left(\PPP\left(V_i (U_L [\Sigma \ 0] U_R)^T \right)^TV_i (U_L [\Sigma \ 0] U_R)^T \right) \\
& \sim & \tr\left(U_L\PPP\left(V_i\left[ \substack{\Sigma \\ 0 } \right]\right)^TV_i \left[ \substack{\Sigma \\ 0 } \right] U_L^T \right)  \\
& \sim & \tr\left(\PPP\left(V_i\left[ \substack{\Sigma \\ 0 } \right]\right)^TV_i \left[ \substack{\Sigma \\ 0 } \right] \right)  \\
&  \sim & \frac1{\sqrt{dp}}\tr\left(\PPP\left(G\sqrt{d}\Sigma \right)^TG \sqrt{d}\Sigma \right),\end{eqnarray*}
where $G$ is a $d\times d$ Gaussian matrix with $\NNN(0,\frac1d)$ entries.

This means that
\[
A_{D}(V) = \frac1n\sum_{i=1}^n X_i,
\]
with $X_i$ i.i.d. distributed as $ \frac1{\sqrt{dp}}\tr\left(\PPP\left(G\sqrt{d}\Sigma \right)^TG \sqrt{d}\Sigma \right)$.

Since $\|\sqrt{d}\Sigma\|_F^2 = d$, by~(\ref{def:alphadstarREAL}), we get
\[
\EE\tr\left(\PPP\left(G\sqrt{d}\Sigma \right)^TG \sqrt{d}\Sigma \right)  \leq  d\alpha_{\RR}^{\ast}(d).
\]
This, together with Theorem~\ref{alphastaralpha}, gives
\begin{equation}\label{eq:wcfinalexpectationcalc}
\EE X_i\leq \sqrt{\frac{d}{p}}\alpha_{\RR}(d).
\end{equation}

In order to give tail bounds for \(A_{D}(V) = \frac1n\sum_{i=1}^n X_i\) we will show that $X_i$ is subgaussian and use Hoeffding's inequality (see Vershynin's notes~\cite{VershyninNARandomMatrices}). In fact, 
\[
X_i \sim \frac1{\sqrt{dp}}\tr\left(\PPP\left(G\sqrt{d}\Sigma \right)^TG \sqrt{d}\Sigma \right) \leq \frac1{\sqrt{p}} \| \PPP\left(G \Sigma \right)  \|_F\| G \Sigma   \|_F = \sqrt{\frac{d}{p}} \| G D   \|_F \leq \sqrt{\frac{d}{p}} \| G\|_F.
\]

Note that $\sqrt{\frac{d}{p}} \| G\|_F$ is a subgaussian random variable as $\|G\|_F$ is smaller than the entry wise $\ell_1$ norm of $G$ which is the sum of $d^2$ half-normals (more specifically, the absolute value of a $\NNN(0,\frac1d)$ random variable). Since half-normals are subgaussian and the sum of subgaussian random variables is a subgaussian random variable with subgaussian norm at most the sum of the norms (see the Rotation invariance Lemma in~\cite{VershyninNARandomMatrices}) we get that $X_i$ is subgaussian. Furthermore, the subgaussian norm of $X_i$, which we define as $\|X_i\|_{\psi_2} = \sup_{p\geq 1} p^{-1/2}(\EE|X|^p)^{1/p}$, is bounded by $\|X_i\|_{\psi_2} \leq C\sqrt{\frac{d^2}p} = C\frac{d}{\sqrt{p}}$, for some universal constant $C$.


Hence, we can use Hoeffding's inequality (see~\cite{VershyninNARandomMatrices}) and get, since $\EE X_i  \leq \sqrt{\frac{d}{p}}\alpha_{\RR}(d)$,
\[
\Prob\left[ A_{D} \geq \sqrt{\frac{d}{p}}\alpha_{\RR}(d) + t \right] \leq \Prob\left[ \left|A_{D} - \EE X_i \right| \geq t \right] \leq \exp\left( 1 - \frac{c_2t^2n}{\|X_i\|_{\psi_2}^2}\right) \leq 3\exp\left( - c_1\frac{t^2p}{d^2}n\right),
\]
where $c_i$ are universal constants.
 
To find an upper bound for $A = \max_{D\in\RR^{d\times pd}: \|D\|_F = 1}A_{D}$ we use a classicl $\epsilon$-net argument. 
There exists a set $\NNN$ of matrices  $D_k\in\RR^{d\times pd}$ satisfying  $\|D_k\|_F = 1$, such that for any $D \in \RR^{d\times pd}$ with Frobenius norm $1$, there exists an element $D_k\in \NNN$ such that $||D - D_k||_F \leq \epsilon$. $\NNN$ is called an $\epsilon$-net, and it's known (see~\cite{VershyninNARandomMatrices}) that there exists such a set with size
\[
|\NNN| \leq \left( 1 + \frac2\epsilon \right)^{d^2p}.
\]

By the union-bound, with probability at least
\[
1- |\NNN|\,\Prob\left[ A_{D} \geq \sqrt{\frac{d}{p}}\alpha_{\RR}(d) + t \right] \geq 1 - \left[\left( 1 + \frac2\epsilon \right)^{d^2p} 3\exp\left( - c_1\frac{t^2n}{d^3}\right)\right],
\]
all the $D_k$'s in $\NNN$ satisfy
\[
A_{D_k} \leq \sqrt{\frac{d}{p}}\alpha_{\RR}^\ast(d) + t.
\]

If $D$ is not in $\NNN$, there exists $D_k\in\NNN$ such that $\|D-D_k\|_F\leq \epsilon$. This means that
\begin{eqnarray*}
A_{D} & \leq & \frac1n\sum_{i=1}^n\tr\left(\PPP\left(V_i D^T \right)^TV_i D_\ast^T \right) + \frac1n\sum_{i=1}^n\tr\left(\PPP\left(V_i D^T \right)^TV_i (D_\ast^T-D^T) \right) \\
& \leq & \frac1n\sum_{i=1}^n\tr\left(\PPP\left(V_i D_\ast^T \right)^TV_i D_\ast^T \right) + \frac1n\sum_{i=1}^n\| \PPP\left(V_i D^T \right)^TV_i \|_F \| D_\ast-D \|_F \\
& \leq & \sqrt{\frac{d}{p}}\alpha_{\RR}^\ast(d) + t + \epsilon\left(\frac1n\sum_{i=1}^n\| V_i \|_F \right).
\end{eqnarray*}

We can globally bound $\left(\frac1n\sum_{i=1}^n\| V_i \|_F \right)$ by Hoeffding's inequality as well (see~\cite{VershyninNARandomMatrices}). Using the same argument as above, it is easy to see that $\| V_i \|_F$ has subgaussian norm bounded by $\sqrt{d}$, and an explicit computation shows its mean is $\frac{1}{\sqrt{dp}}\sqrt{2}\frac{\Gamma((d^2p+1)/2)}{\Gamma(d^2p/2)} \leq 2\sqrt{d}$, where the inequality follows from lemma \ref{lemma:alphacbounds}. 

This means that by Hoeffding's inequality (see~\cite{VershyninNARandomMatrices}) 
\begin{eqnarray*}
\Prob\left[ \frac1n\sum_{i=1}^n\| V_i \|_F \geq 2\sqrt{d} + t  \right] &\leq& \exp\left( 1 - \frac{c_4t^2n}{\|\| V_i \|_F\|_{\psi_2}^2}\right) \\
&\leq& 3\exp\left( - c_3\frac{t^2n}{d }\right),
\end{eqnarray*}
with $c_i$ universal constants.

By union-bound on the two events above, with probability at least
\[1 - 3\exp\left( - c_3\frac{t^2n}{d}\right) - \left[\left( 1 + \frac2\epsilon \right)^{d^2p} 3\exp\left( - c_1\frac{t^2n}{d^3}\right)\right],
\]
we have
\[
A \leq \sqrt{\frac{d}{p}}\alpha_{\RR}^\ast(d) + t + \epsilon(2d+t).
\]

Choosing $t = \frac1{2p}$ and $\epsilon = \frac1{6dp}$ we get
\[
A \leq \sqrt{\frac{d}{p}}\alpha_{\RR}^\ast(d) + \frac1p,
\]

with probability at least
\begin{eqnarray*}
& & 1 - 3\exp\left( - c_3\frac{n}{4p^2}\right) - \left[\left( 1 + 12dp \right)^{d^2p} 3\exp\left( - c_1\frac{n}{4p^2d^3}\right)\right] \\ 
& = & 1 - 6\left[\left( 1 + 12dp \right)^{d^2p} 3\exp\left( - c_1\frac{n}{4p^2d^3}\right)\right]\\ 
\end{eqnarray*}
which can be made arbitrarily close to $1$ by taking $n$ large enough.

This means that
\[
\max_{O_1,.. .,O_n \in \OOO_d}\left\|\frac1n\sum_{i=1}^n O_i^T V_i   \right\|_F^2 \leq \frac{d}p\alpha_{\RR}(d)^2 + \frac1p,
\]
with high probability, proving the lemma.
$\qed$
}

Regarding $w_r$, we know that it is at least the value of $\frac1{n^2}\sum_{i,j}^n\tr\left((V_iV_j^T)^TX_iX_j^T\right)$ for $X_i = \PPP(V_i)$. Since, for $p$ large enough, $V_iV_i^T \approx I_{d\times d}$ we essentially have $w_r \gtrsim \frac1{n^2}\sum_{i,j}^n\|V_iV_j^T\|_F^2$ which should approximate $\EE \|V_iV_j^T\|_F^2 \approx \frac{d}p$. This is made precise in the following lemma:
\begin{lemma}
For any $d$ and $\varepsilon>0$ there exists $p_0$ and $n_0$ such that, for any $p>p_0$ and $n>n_0$,
\[
 \frac1{n^2}\sum_{i,j}^n\tr\left((V_iV_j^T)^T\PPP_{(d,dp)}(V_i)\PPP_{(d,dp)}(V_j)^T\right) \geq \frac{d}p - \varepsilon,
\]
with probability strictly larger than $1/2$.
\end{lemma}
\proof{
Recall that $\PPP_{(d,dp)}(V_i)$ is the $d\times dp$ matrix polar component of $V_i$, meaning that
$$\tr\left( \PPP_{(d,dp)}(V_i)^T V_i \right) = \sum_{k=1}^d\sigma_k(V_i).$$
Hence,
\begin{eqnarray*}
\frac1{n^2}\sum_{i,j}^n\tr\left((V_iV_j^T)^T\PPP_{(d,dp)}(V_i)\PPP_{(d,dp)}(V_j)^T\right)  & = & \left\| \frac1n \sum_{i=1}^n \PPP(V_i)^T V_i \right\|_F^2 \\
& \geq & \frac1{\left\| I_{dp\times dp}  \right\|_F^2}\left[ \frac1n \sum_{i=1}^n\tr\left( \PPP(V_i)^T V_i I_{dp\times dp} \right) \right]^2\\
& = & \frac1{dp}\left[ \frac1n \sum_{i=1}^n\sum_{k=1}^d\sigma_k(V_i) \right]^2.
\end{eqnarray*}

We proceed by using a lower bound for the expected value of the smallest eigenvalue (see~\cite{VershyninNARandomMatrices}), and get
\[
\EE \sum_{k=1}^d\sigma_k(V_i) \geq d\ \EE \sigma_{\min}(V_i) = d \left( 1 -\frac1{\sqrt{p}} \right).
\] 

Since $\sum_{k=1}^d\sigma_k(V_i) \leq \sqrt{d}\|V_i\|_F$, it has subgaussian norm smaller than $Cd$, with $C$ an universal constant (using the same argument as in Lemma \ref{lemma:wcnotappendix}). Therefore, by Hoeffding's inequality (see~\cite{VershyninNARandomMatrices}),
\begin{eqnarray*}
\Prob\left[  \frac1n \sum_{i=1}^n\sum_{k=1}^d\sigma_k(V_i)  \leq  d \left( 1 -\frac1{\sqrt{p}}\right) - t \right] 
& \leq & \exp\left(1 - \frac{c_1t^2}{\left\|\sum_{k=1}^d\sigma_k(V_i)\right\|_{\psi_2}^2  }n \right) \\
& \leq & \exp\left(1 - c_2\frac{t^2}{d^2}n \right),
\end{eqnarray*}
where $c_i$ are universal constants.

By setting $t = \frac{d}{\sqrt{p}}$, we get
\[
\frac1{n^2}\sum_{i,j}^n\tr\left((V_iV_j^T)^T\PPP_{(d,dp)}(V_i)\PPP_{(d,dp)}(V_j)^T\right)  \geq \frac{d}{p}\left( 1 - 2\frac1{\sqrt{p}} \right)^2,
\]
with probability at least $1 - \exp\left(1 - c_2\frac{1}{p}n \right) = 1-o_n(1)$, proving the Lemma.
$\qed$
}

Theorem~\ref{theoremtightnessratios} immediately follows from these two lemmas.

We note that these techniques are quite general. It is not difficult to see that these arguments, establishing integrality gaps that match the approximation ratios obtained, can be easily adapted for both the unitary case and the rank constrained case introduced in Section~\ref{Section:HigherRank}. For the sake of exposition we omit the details in these cases.

\section{Open Problems and Future Work}\label{section:futurework}

Besides Conjecture \ref{conjecture:alphar}, there are several extensions of this work that the authors consider to be interesting directions for future work.

A natural extension is to consider the little Grothendieck problem (\ref{eq_QFOG}) over other groups of matrices. One interesting extension would be to consider the special orthogonal group $S\OOO_d$ and the special unitary group $S\UUU_d$, these seem more difficult since they are not described by quadratic constraints.~\footnote{The additional constraint that forces a matrix to be in the special orthogonal or unitary group is having determinant equal to $1$ which is not quadratic.}

In some applications, like Synchronization~\cite{Bandeira_Singer_Spielman_OdCheeger,ASinger_2011_angsync} (a similar problem to Orthogonal Procrustes) and Common Lines~\cite{ASinger_YShkolnisky_commonlines}, the positive semidefiniteness condition is not natural. It would be useful to better understand approximation algorithms for a version of~(\ref{eq_QFOG}) where $C$ is not assumed to be positive semidefinite. Previous work in the special case $d=1$, \cite{Nemirovski_etal_losinglogEllipsoids,Charikar_littleGrothendieck,Alon_etal_generalGrothendieck} for $\OOO_1$ and~\cite{So_Zhang_Ye_QuadSDPrelax} for $\UUU_1$, suggest that it is possible to obtain an approximation ratio for (\ref{eq_QFOG}) depending logarithmically on the size of the problem. Moreover, for $\OOO_1$, the logarithmic term is known to be needed in general~\cite{Alon_etal_generalGrothendieck}.

It would also be interesting to understand whether the techniques in \cite{NAlon_ANaor_2006} can be adapted to obtain an approximation algorithm to the bipartite Grothendieck problem over the orthogonal group; this would be closer in spirit to the non commutative Grothendieck inequality~\cite{Naor_etal_NCGI}.

Another interesting question is whether the approximation ratios obtained in this paper correspond to the hardness of approximation of the problem (perhaps conditioned on the Unique-Games conjecture~\cite{SKhot_2010}). Our optimality conditions are restricted to the particular relaxation we consider and do not exclude the existence of an efficient algorithm, not relying on the same relaxation, that approximates (\ref{eq_QFOG}) with a better approximation ratio. Nevertheless, Raghavendra~\cite{Raghavendra_tightratiosUG} results on hardness for a host of problems matching the integrality gap of natural SDP relaxations suggest that our approximation ratios might be optimal (see also the recent results in~\cite{Briet_GrothendieckHardness}).


\section*{Acknowledgments}
The authors would like to thank Moses Charikar for valuable guidance in context of this work and  Jop Briet, Alexander Iriza, Yuehaw Khoo, Dustin Mixon, Oded Regev, and Zhizhen Zhao for 
insightful discussions on the topic of this paper. Special thanks to Johannes Trost for a very useful answer to a Mathoverflow question posed by the first author. Finally, we would like to thank the reviewers for numerous suggestions that helped to greatly improve the quality of this paper.

A.~S.~Bandeira was supported by AFOSR Grant No. FA9550-12-1-0317. A. Singer was partially supported by Award Number FA9550-12-1-0317 and FA9550-13-1-0076 from
AFOSR, by Award Number R01GM090200 from the NIGMS, and by Award Number LTR DTD 06-05-2012 from the Simons Foundation. Parts of this work have appeared in C. Kennedy's senior thesis at Princeton University.

\bibliographystyle{alpha}



\bibliography{OrthogonalCut_bib}


\appendix

\section{Technical proofs - analysis of algorithm for the Stiefel Manifold setting}

\begin{lemma}\label{lemma:alphad}
Let $r\geq d$.
Let $G$ be a $d\times r$ Gaussian random matrix with real valued i.i.d. $\NNN\left(0,\frac1r\right)$ entries and let $\alpha_{\RR}(d,r)$ as defined in Definition~\ref{def:alphadr}. Then, 
\[
\EE \left(\PPP_{d,r}(G) G^T\right)= \EE \left(G\PPP_{d,r}(G)^T\right)= \alpha_{\RR}(d,r)I_{d\times d}.
\]
Furthermore, if $G$ is a $d\times r$ Gaussian random matrix with complex valued i.i.d. $\NNN\left(0,\frac1r\right)$ entries and $\alpha_{\CC}(d,r)$ the analogous constant (Definition~\ref{def:alphadr}), then
\[
\EE \left(\PPP_{d,r}(G) G^H\right)= \EE \left(G\PPP_{d,r}(G)^H\right)= \alpha_{\CC}(d,r)I_{d\times d}.
\]
\end{lemma}

The proof of this Lemma is a simple adaptation of the proof of Lemma~\ref{lemma:alphad_dequalr}.

\proof{
We restrict the presentation to the real case.  As before, all the arguments are equivalent to the complex case, replacing all transposes with Hermitian adjoints and $\alpha_{\RR}(d,r)$ with $\alpha_{\CC}(d,r)$.

Let $G = U [\Sigma \ 0] V^T$ be the singular value decomposition of $G$. Since $GG^T = U\Sigma^2 U^T$ is a Wishart matrix, it is well known that its eigenvalues and eigenvectors are independent and $U$ is distributed according to the Haar measure in $\OOO_d$ (see e.g. Lemma 2.6 in~\cite{Tulino_Verdu_RM}). To resolve ambiguities, we consider $\Sigma$ ordered such that $\Sigma_{11} \geq \Sigma_{22} \geq.. .\geq\Sigma_{dd}$.

Let $Y = \PPP_{(d,r)}(G) G^T$. Since
\[
\PPP_{(d,r)}(G) = \PPP_{(d,r)}(U [\Sigma \ 0] V^T) = U[I_{d\times d} \ 0]V^T,
\]
we have
\[
Y = \PPP_{(d,r)}(U [\Sigma \ 0]V^T) (U [\Sigma \ 0] V^T)^T = U [I_{d\times d} \ 0]V^T V \Sigma U^T = U \Sigma U^T.
\]
Note that $ G\PPP_{(d,r)}(G)^T = U \Sigma U^T =  Y$.

Since $Y_{ij} = u_i \Sigma u_j^T$, where $u_1,\dots,u_d$ are the rows of $U$, and $U$ is distributed according to the Haar measure, we have that $u_j$ and $-u_j$ have the same distribution conditioned on any $u_i$, for $i\neq j$, and $\Sigma$. This implies that, if $i\neq j$, $Y_{ij} = u_i \Sigma u_j^T$ is a symmetric random variable, and so $\EE Y_{ij} = 0$. Also, $u_i\sim u_j$ implies that $Y_{ii}\sim Y_{jj}$. This means that $\EE Y = c I_{d\times d}$ for some constant $c$. To obtain $c$,
\[
c = c\frac1d\tr(I_{d\times d}) = \frac1d\EE \tr(Y) = \frac1d\EE \tr(U  \Sigma U^T) = \frac1d\EE\sum_{k=1}^n\sigma_{k}(G) = \alpha_{\RR}(d,r),
\]
which shows the lemma.
$\qed$}

\begin{lemma}\label{lemma_polar1}
Let $r\geq d$. Let $M,N\in \RR^{d\times nd}$ such that $MM^T = NN^T = I_{d \times d}$. Let $R \in \RR^{nd \times r}$ be a Gaussian random matrix with real valued i.i.d entries $\NNN\left( 0, \frac1r \right)$. Then
\[
 \EE\left[\PPP_{(d,r)}(M R) (N R)^T\right] =  \EE\left[(M R) \PPP_{(d,r)}(N R)^T\right] = \alpha_{\RR}(d,r) MN^T,
\]
where $\alpha_{\RR}(d,r)$ is the constant in Definition \ref{def:alphadr}.

Analogously, if  $M,N\in \CC^{d\times nd}$ such that $MM^H = NN^H = I_{d \times d}$, and $R \in \CC^{nd \times r}$ is a Gaussian random matrix with complex valued i.i.d entries $\NNN\left( 0, \frac1r \right)$, then
\[
 \EE\left[ \PPP_{(d,r)}(M R) (N R)^H\right] = \EE\left[ (M R) \PPP_{(d,r)}(N R)^H\right] = \alpha_{\CC}(d,r)   MN^H,
\]
where $\alpha_{\CC}(d,r)$ is the constant in Definition \ref{def:alphadr}.

\end{lemma}

Similarly to above, the proof of this Lemma is a simple adaptation of the proof of Lemma~\ref{lemma_polar1_dequalr}.

\proof{
We restrict the presentation of proof to the real case.  Nevertheless, all the arguments trivially adapt to the complex case by, essentially, replacing all transposes with Hermitian adjoints and $\alpha_{\RR}(d)$ and $\alpha_{\RR}(d,r)$ with $\alpha_{\CC}(d)$ and $\alpha_{\CC}(d,r)$.

Let $A = \left[ M^T\text{  } N^T\right]\in\RR^{dn\times 2d}$ and $A=QB$ be the $QR$ decomposition of $A$ with  $Q\in \RR^{nd\times nd}$ an orthogonal matrix and $B \in \RR^{nd \times 2d}$ upper triangular with non-negative diagonal entries; note that only the first $2d$ rows of $B$ are nonzero. We can write
\[
Q^TA = B = \begin{bmatrix} B_{11} & B_{12}\\
0_d & B_{22}\\
0_d & 0_d\\\
\vdots & \vdots\\
0_d & 0_d
\end{bmatrix} \in \RR^{dn \times 2d},
\]
where $B_{11}\in\RR^{d\times d}$ and $B_{22}\in\RR^{d\times d}$ are upper triangular matrices with non-negative diagonal entries. Since $$(Q^T M^T)^T_{11}(Q^T M^T)_{11} = (Q^T M^T)^T(Q^T M^T) = M Q Q^T M^T = M I_{nd \times nd} M^T = M M^T = I_{d \times d},$$ $B_{11} = (Q^T M^T)_{11}$ is an orthogonal matrix, which together with the non-negativity of the diagonal entries (and the fact that $B_{11}$ is upper-triangular) forces $B_{11}$ to be $B_{11} = I_{d\times d}$.

Since $R$ is a Gaussian matrix and $Q$ is an orthogonal matrix, $QR \sim R$ which implies
\[
 \EE\left[ \PPP_{(d,r)}(M R) (N R)^T\right] = \EE \left[ \PPP_{(d,r)}(M Q R)(NQ R)^T \right].
\]
Since $MQ=[B_{11}^T, 0_{d\times d},\cdots,0_{d\times d}] = [I_{d\times d}, 0_{d\times d},\cdots,0_{d\times d}]$ and $NQ = [B_{12}^T, B_{22}^T, 0_{d\times d},\cdots,0_{d\times d}]$,
\[
\EE\left[ \PPP_{(d,r)}(M R) (N R)^T\right] =  \e \left[\PPP_{(d,r)}(R_1)(B_{12}^T R_1 + B_{22}^T R_2)^T\right],
\]
where $R_1$ and $R_2$ are the first two $d\times r$ blocks of $R$. Since these blocks are independent, the second term vanishes and we have
\[
\EE\left[ \PPP_{(d,r)}(M R) (N R)^T\right] =  \EE\left[ \PPP_{(d,r)}(R_1) R_1^T\right]B_{12}.
\]
The Lemma now follows from using Lemma \ref{lemma:alphad} to obtain $\EE \left[\PPP_{(d,r)}(R_1) R_1^ T\right]= \alpha_{\RR}(d,r)I_{d\times d}$ and noting that $B_{12} = (Q^TM^T)^T(Q^TN^T) = MN^T$.

The same argument, with $Q'B'$ the $QR$ decomposition of $A' = \left[ N^T M^T\right]\in\RR^{dn\times 2d}$ instead, shows
\[
\EE\left[ (M R) \PPP_{(d,r)}(N R)^T\right] = \EE \left[  R_1\PPP_{(d,r)}(R_1)^T\right] MN^T = \alpha_{\RR}(d,r)MN^T.
\]

$\qed$}

\section{Bounds for the average singular value}\label{sec:appendixbounds}

\begin{lemma}\label{lemma:alphacbounds}
Let $G_{\CC} \in \CC^{d\times d}$ be a Gaussian random matrix with i.i.d. complex valued $\mathcal{N}(0,d^{-1})$ entries and define $\alpha_{\CC}(d):= \EE\left[ \frac{1}{d} \sum_{j=1}^d \sigma_j(G_{\CC})\right]$.  We have the following bound
\[
\alpha_{\CC}(d) \geq \frac8{3\pi} - \frac{5.05}d.
\]
\end{lemma}
\proof{
We express $\alpha_{\CC}(d)$ as sums and products of Gamma functions and then use classical bounds to obtain our result.\\

Recall that from equation (\ref{sum:alphaC}),
\begin{equation} \label{equation:alphaC}
\alpha_{\CC}(d) = d^{-3/2} \sum_{n=0}^{d-1} T_n,
\end{equation}
where 
\[
T_n = \int_0^{\infty} x^{1/2} e^{-x} L_n(x)^2 dx,
\]
and $L_n(x)$ is the $n$th Laguerre polynomial,
\[
L_n(x) = \sum_{k=0}^n \binom{n}{k} \frac{(-1)^k}{k!} x^k.
\]
This integral can be expressed as (see \cite{Gradshteyn_Ryzhik} section 7.414 equation 4(1))
\begin{equation} \label{equation:Tn}
T_n = \frac{ \Gamma(n+3/2)}{\Gamma(n+1)} \sum_{m=0}^n \frac{\left(\frac{-1}{2}\right)_m  (-n)_m}{(m!)^2 (-n-\frac{1}{2})_m},
\end{equation}
where $(x)_m$ is the Pochhammer symbol
\[
(x)_m = \frac{\Gamma(x+m)}{\Gamma(x)}.
\]
The next lemma states a couple basic facts about the Gamma function that we will need in the subsequent computations.
\begin{lemma} \label{lemma:alphacbounds}
The Gamma function satisfies the following inequalities:
\[
\frac{1}{\sqrt{n}}  \leq \frac{\Gamma(n)}{\Gamma(n+1/2)} \leq \frac{1}{\sqrt{n-1/2}}
\]
\[
\sqrt{n} \leq \frac{\Gamma(n+1)}{\Gamma(n+1/2)} \leq \sqrt{n+1/2}.
\]
\end{lemma}
\proof{ See \cite{abramowitz_stegun} page 255.
}
\\
\\
We want to bound the summation in (\ref{equation:Tn}), which we rewrite as
\[
\sum_{m=0}^n \frac{\left(\frac{-1}{2}\right)_m  (-n)_m}{(m!)^2 (-n-\frac{1}{2})_m} = \sum_{m=0}^{\infty} \frac{(\frac{-1}{2})_m^2}{(m!)^2} - \sum_{m= n+1}^{\infty}\frac{(\frac{-1}{2})_m^2}{(m!)^2} -\sum_{m=0}^n \frac{(\frac{-1}{2})_m^2}{(m!)^2} \left(1 - \frac{(-n)_m}{(-n-\frac{1}{2})_m} \right).
\]
For simplicity define
\[
(I) := \sum_{m=0}^{\infty} \frac{(\frac{-1}{2})^2_m}{(m!)^2}
\]
\[
(II) :=  \sum_{m= n+1}^{\infty}\frac{(\frac{-1}{2})_m^2}{(m!)^2}
\]
\[
(III) := \sum_{m=0}^n \frac{(\frac{-1}{2})_m^2}{(m!)^2} \left(1 - \frac{(-n)_m}{(-n-\frac{1}{2})_m} \right),
\]
so that (\ref{equation:Tn}) becomes
\[
T_n = \frac{\Gamma(n+3/2)}{\Gamma(n+1)} ((I) + (II) + (III)).
\]

The first term we can compute explicitly (see \cite{Gradshteyn_Ryzhik}) as
\[
(I) = \frac{4}{\pi}.
\]
For the second term we use the fact that $(\frac{-1}{2})_m = \Gamma(m-1/2)/\Gamma(-1/2)$ to get
\[
(II)= \sum_{m = n+1}^{\infty} \frac{1}{\Gamma(-1/2)^2} \frac{\Gamma(m-1/2)^2}{\Gamma(m+1)^2} = \frac{1}{4\pi} \sum_{m=n+1}^{\infty} \frac{\Gamma(m-1/2)^2}{\Gamma(m+1)^2}.
\]
Using the first inequality in Lemma \ref{lemma:alphacbounds} and the multiplication formula for the Gamma function,
\[
\frac{\Gamma(m-1/2)}{\Gamma(m+1)} = \frac{1}{m-1/2} \frac{\Gamma(m+1/2)}{\Gamma(m+1)} \leq \frac{1}{(m-1/2)\sqrt{m}}
\]
so we have
\[
(II) \leq \frac{1}{4\pi} \sum_{m=n+1}^{\infty} \frac{1}{(m-1/2)^2 m} \leq \frac{1}{4\pi} \int_{n-1/2}^{\infty} \frac{1}{x^3} dx = \frac{1}{2\pi (2n-1)^2}.
\]
For the third term, we use the formula $(x)_m = \frac{\Gamma(x+n)}{\Gamma(x)}$ to deduce
\begin{align*}
(III) &= \sum_{m=0}^n \frac{(\frac{-1}{2})_m^2}{(m!)^2} \left(1 - \frac{(-n)_m}{(-n-\frac{1}{2})_m} \right)\\
&= \frac{1}{4\pi} \sum_{m=0}^n \frac{\Gamma(m-1/2)^2}{\Gamma(m+1)^2} \left(1 - \frac{\Gamma(n+1) \Gamma(n-m+3/2)}{\Gamma(n+3/2) \Gamma(n-m+1)}\right)\notag\\
&= \frac{\Gamma(n+1)}{\Gamma(n+3/2)} \frac{1}{4\pi} \sum_{m=0}^{n} \frac{\Gamma(m-1/2)^2}{\Gamma(m+1)^2} \left(\frac{\Gamma(n+3/2)}{\Gamma(n+1)} - \frac{\Gamma(n-m+3/2)}{\Gamma(n-m+1)}\right).
\end{align*}
Using the second bound in Lemma \ref{lemma:alphacbounds},
\[
\frac{\Gamma(n-m+3/2)}{\Gamma(n-m+1)} \geq \sqrt{n-m+1/2},
\]
and also
\[
\frac{\Gamma(n+3/2)}{\Gamma(n+1)} \leq \sqrt{n+1},
\]
so that
\begin{align*}
(III) \leq \frac{1}{4\pi \sqrt{n+1/2}} \sum_{m=0}^n \left(\frac{1}{(m-1/2)\sqrt{m+1/2}}\right)^2 \left(\sqrt{n+1} - \sqrt{n-m+1/2}\right).
\end{align*}
If we multiply top and bottom by $\sqrt{n+1} + \sqrt{n-m+1/2}$ and use the fact that
\[
\frac{m+1/2}{\sqrt{n+1} + \sqrt{n-m+1/2}} \leq \frac{m+1/2}{\sqrt{n+1}},
\]
then
\begin{align*}
(III) &\leq \frac{1}{4\pi \sqrt{n+1/2}} \sum_{m=0}^n \frac{1}{(m-1/2)^2 (m+1/2)} \frac{m+1/2}{\sqrt{n+1}}\\
& \leq \frac{1}{2\pi(n+1)} \sum_{m=0}^n \frac{1}{(m-1/2)^2}\\
& \leq \frac{1}{n+1} \frac{8+\pi^2}{2\pi}\\
&\leq  \frac{3}{n+1}.
\end{align*}
Combining our bounds for (I), (II) and (III),
\begin{align*}
T_n &= \frac{\Gamma(n+3/2)}{\Gamma(n+1)} \left[ (I) - (II)-(III)\right]\\
&\geq \frac{\Gamma(n+3/2)}{\Gamma(n+1)}\left (\frac{4}{\pi} - \frac{1}{2\pi(2n-1)^2} - \frac{3}{n+1}\right)\\
&\geq \sqrt{n+1/2} \left (\frac{4}{\pi} - \frac{1}{2\pi(2n-1)^2} - \frac{3}{n+1}\right),
\end{align*}
and by (\ref{equation:alphaC}),
\[
\alpha_{\CC}(d) \geq \frac{1}{d^{3/2}} \sum_{n=1}^{d-1} \sqrt{n+1/2} \left (\frac{4}{\pi} - \frac{1}{2\pi(2n-1)^2} - \frac{3}{n+1}\right).
\]
The term $\frac{1}{d^{3/2}} \sum_{n=1}^{d-1} 4 \sqrt{n+1/2}/ \pi $ is the main term and can be bounded below  by
\begin{align*}
 \frac{1}{d^{3/2}} \sum_{n=1}^{d-1} \frac{4 \sqrt{n+1/2}}{ \pi} &\geq \frac{1}{d^{3/2}} \frac{8}{3\pi} \left((d-1/2)^{3/2} - (1/2)^{3/2}\right) \\& \geq (1 - (2d)^{-1}) \frac{8}{3\pi}  - (2d)^{-3/2}\\
& \geq \frac{8}{3\pi} - \left(\frac{8}{3\pi} + \frac{1}{2}\right)d^{-1} .
\end{align*}\\
\\
The other error terms are at most
\begin{align*}
 d^{-3/2} \sum_{n=1}^{d-1} \sqrt{n+1/2}\left(\frac{1}{2\pi (2n-1)^2} + \frac{3}{n+1}\right) &\leq \frac{1}{d^{3/2}} \sum_{n=1}^{d-1}\frac{4}{\pi(n+1)} \sqrt{n+1/2}\\
&\leq \frac{1}{d^{3/2}} \sum_{n=1}^{d-1} \frac{4}{\pi(n+1)^{1/2}}\\
&\leq \frac{4}{\pi d^{3/2}}2\sqrt{d+1}.
\end{align*}
Combining the main and error term bounds, the lemma follows.
$\qed$
}
\begin{lemma}\label{lemma:alpharcdiff}
For $G_{\KK} \in \KK^{d\times d}$ a Gaussian  random matrix with i.i.d. $\KK$ valued $\mathcal{N}(0,d^{-1})$ entries, define $\alpha_{\KK}(d) := \EE\left[ \frac{1}{d} \sum_{j=1}^d \sigma_j(G_{\KK})\right]$.  The following holds
\[
\alpha_{\CC}(d) - \alpha_{\RR}(d) \leq 4.02 d^{-1}.
\]
\end{lemma}
\proof{\\

To find an explicit formula for $\alpha_{\RR}(d)$, we need an expression for the spectral distribution of the wishart matrix $d G_{\RR} G_{\RR}^T$, which we call $p_d^{\RR}(x)$, given by equation (16) in \cite{Livan11_WishartLaguerre}:
\[
p_d^{\RR}(x) = \frac1{2d}\left( 2R_d(x) - \frac{\Gamma\left( \frac{d}{2}+\frac12 \right)}{\Gamma\left( \frac{d}{2} \right)}L_{d-1}(x) \left\{ \psi_1\left(x\right) - \psi_2\left(x\right)  \right\} \right), 
\]
where
\[
\psi_1(x) = e^{-x} \sum_{k=0}^{(\kappa+d-2)/2} \delta_k L_{2k+1-\kappa}(x),
\]
\[
\psi_2(x) = \left(\frac{x}2\right)^{-1/2}e^{-\frac{x}2} \left[ (1-\kappa)\frac{2\Gamma\left( \frac12,\frac{x}2 \right)}{\Gamma\left( \frac12\right)}   + 2\kappa -1  \right],
\]
\[
R_d(x) = e^{-x}\sum_{m=0}^{d-1}\left(L_{m}(x) \right)^2,
\]
\[
\delta_k = \frac{\Gamma\left( k + 1 -  \frac\kappa2 \right)}{\Gamma\left( k + \frac32 -  \frac\kappa2\right)},
\]
$\kappa = d\mod 2$ and $\Gamma(a,y) = \int_y^\infty t^{a-1}e^{-t}dt$ is the incomplete Gamma function.\\
\\
This means that
\begin{align*}
  \alpha_{\RR}(d) &=  d^{-1/2} \int_0^{\infty} x^{1/2} p_d^{\RR}(x) dx\\
&= \frac1{d^{3/2}}\int_0^\infty x^{1/2}R_d(x)dx - \frac1{2d^{3/2}}\int_0^\infty x^{1/2} \frac{\Gamma\left( \frac{d}{2}+\frac12 \right)}{\Gamma\left( \frac{d}{2} \right)}L_{d-1}(x) \left\{ \psi_1\left(x\right) - \psi_2\left(x\right)  \right\} dx.
\end{align*}
Recall that (see section 5)
\[
\alpha_{\CC}(d) = d^{-3/2}\sum_{n=0}^{d-1}\int_{0}^\infty x^{1/2}e^{-x}L_n(x)^2 dx
\]
which implies
 \begin{equation}\label{alpharcdiff}
  \alpha_{\RR}(d) =  \alpha_{\CC}(d) - \frac1{2d^{3/2}}\int_0^\infty  x^{1/2}\frac{\Gamma\left( \frac{d}{2}+\frac12 \right)}{\Gamma\left( \frac{d}{2} \right)}L_{d-1}(x) \left\{ \psi_1\left(x\right) - \psi_2\left(x\right)  \right\} dx.
 \end{equation}
 We are especially interested in the following terms which appear in the full expression for $\alpha_{\RR}(d)$:
 \begin{equation}\label{summationq}
 Q(m,k)= \int_0^{\infty} x^{1/2} e^{-x} L_m(x) L_k(x) dx.
 \end{equation}
From \cite{Gradshteyn_Ryzhik} section 7.414 equation 4(1), we have
 \[
 Q(m,k) = \frac1{4\pi} \sum_{i=0}^{\min\{m,k\}} \frac{\Gamma(i+3/2)}{\Gamma(i+1)}\frac{\Gamma(m-i-1/2)}{\Gamma(m-i+1)}\frac{\Gamma(k-i-1/2)}{\Gamma(k-i+1)}.
 \]
The following lemma deals with bounds on sums involving $Q(m,k)$ terms.
\begin{lemma}\label{lemma:qbounds}
For $Q(m,k)$ as defined in (\ref{summationq}) we have the following bounds
\begin{equation}\label{oddqbound}
\sum_{k=0}^m \frac{\Gamma(k+1/2)}{\Gamma(k+1)} Q(2m,2k) \leq 2.8
\end{equation}
\begin{equation}\label{evenqbound}
\sum_{k=1}^m \frac{\Gamma(k+3/2)}{\Gamma(k+1)} Q(2m-1,2k-1) \leq 5.6
\end{equation}
\end{lemma}
\proof{
Note that in (\ref{oddqbound}),
\[
Q(2m,2k) = \frac{1}{4\pi} \sum_{i=0}^{2k} \frac{\Gamma(i+3/2)}{\Gamma(i+1)} \frac{ \Gamma(2m - i - 1/2)}{\Gamma(2m-i+1)}\frac{ \Gamma(2k - i - 1/2)}{  \Gamma(2k-i+1)}
\]
since $m\geq k$.\\
For $0<i<2k-1$, the $i$th term in the summation of $Q(2m,2k)$ can be bounded above by
\begin{align*}
\frac{\Gamma(i+3/2)}{\Gamma(i+1)} \frac{ \Gamma(2m - i - 1/2)}{\Gamma(2m-i+1)}\frac{ \Gamma(2k - i - 1/2)}{  \Gamma(2k-i+1)} &\leq \sqrt{i+1} \frac{1}{(2k-i)\sqrt{2k-i-1}}\frac{1}{(2m-i)\sqrt{2m-i-1}}\\
 &\leq \sqrt{i+1} \frac{1}{(2k-i-1)^{3/2} (2m-i-1)^{3/2}}.
\end{align*}
This means that
\begin{align*}
Q(2m,2k) &\leq\frac{1}{8 \sqrt{\pi}} \frac{\Gamma(2m-1/2) \Gamma(2k-1/2)}{\Gamma(2m+1) \Gamma(2k+1)}\\
&+ \frac{1}{4\pi} \sum_{i=1}^{2k-1} \sqrt{i+1} \frac{1}{(2k-i-1)^{3/2} (2m-i+1)^{3/2}}\\
&+ \frac{1}{4\pi} \sqrt{\pi} \frac{\Gamma(2k+1/2)}{\Gamma(2k)} \frac{\Gamma(2m-2k+1/2)}{\Gamma(2m-2k+2)}\\
&+ \frac{1}{4\pi} \max \left( \frac{\Gamma(2k+3/2)}{\Gamma(2k+1)} \frac{\Gamma(2m-2k-1/2)}{\Gamma(2m-2k+1)}(-2\sqrt{\pi}) , 0\right).
\end{align*}
We bound the sum from $i=1$ to $2k-3$ by
\begin{align*}
\frac{1}{4\pi} \sum_{i=1}^{2k-3} \sqrt{i+1} \frac{1}{(2k-i-1)^{3/2} (2m-i+1)^{3/2}} &\leq \frac{1}{4\pi} \sum_{i=0}^{2k-3} \sqrt{i+1} \frac{1}{(2k-i-1)^{3/2}} \frac{1}{(2m-2k+1)^{3/2}}\\
& \leq \frac{1}{4\pi (2m-2k+1)^{3/2}} \int_0^{2k-2} \frac{\sqrt{x+1}}{(2k-x-1)^{3/2}} dx\\
&\leq \frac{1}{4\pi(2m-2k+1)^{3/2}} \left(\sqrt{8k} + \frac{\sqrt{4k}}{2k-1}\right),
\end{align*}
so that for $k \geq 1$,
\begin{align*}
Q(2m,2k) &\leq \frac{1}{8\sqrt{\pi}} \frac{\Gamma(2m-1/2) \Gamma(2k-1/2)}{\Gamma(2m+1) \Gamma(2k+1)}\\
&+ \frac{1}{4\pi (2m-2k+1)^{3/2}} \left(\sqrt{8k} + \frac{\sqrt{4k}}{2k-1}\right)\\
&+ \frac{1}{4\pi} \frac{\sqrt{2k-1}}{(2m-2k+3)^{3/2}}\\
&+ \frac{1}{4\pi} \sqrt{\pi} \frac{\Gamma(2k+1/2)}{\Gamma(2k)} \frac{\Gamma(2m-2k+1/2)}{\Gamma(2m-2k+2)}\\
&+ \frac{1}{4\pi} \max \left( \frac{\Gamma(2k+3/2)}{\Gamma(2k+1)} \frac{\Gamma(2m-2k-1/2)}{\Gamma(2m-2k+1)}(-2\sqrt{\pi}) , 0\right).
\end{align*}
For $k=0$, $Q(2m,0) < 0$ except for the term $Q(0,0) = \sqrt{\pi}/2$ which also becomes negative in the full sum, so we ignore these terms.\\
We now turn our attention to the full sum $\sum_{k=0}^m \frac{\Gamma(k+1/2)}{\Gamma(k+1)} Q(2m,2k)$.  As before, we define for clarity
\begin{align*}
(I)&:=\frac{1}{8\sqrt{\pi}}  \sum_{k=1}^m  \frac{\Gamma(k+1/2)}{\Gamma(k+1)}  \frac{\Gamma(2m-1/2) \Gamma(2k-1/2)}{\Gamma(2m+1) \Gamma(2k+1)}\\
(II)&:=\frac{1}{4\pi} \sum_{k=1}^m  \frac{\Gamma(k+1/2)}{\Gamma(k+1)} \frac{1}{(2m-2k+1)^{3/2}} \left(\sqrt{8k} + \frac{\sqrt{4k}}{2k-1}\right)\\
(III)&:=\frac{1}{4\pi} \sum_{k=1}^m  \frac{\Gamma(k+1/2)}{\Gamma(k+1)} \frac{\sqrt{2k-1}}{(2m-2k+3)^{3/2}}\\
(IV)&:= \frac{1}{4\sqrt{\pi} } \sum_{k=1}^m  \frac{\Gamma(k+1/2)}{\Gamma(k+1)} \frac{\Gamma(2k+1/2)}{\Gamma(2k)} \frac{\Gamma(2m-2k+1/2)}{\Gamma(2m-2k+2)}\\
(V)& := \frac{1}{4\pi} \sum_{k=1}^m  \frac{\Gamma(k+1/2)}{\Gamma(k+1)} \max \left( \frac{\Gamma(2k+3/2)}{\Gamma(2k+1)} \frac{\Gamma(2m-2k-1/2)}{\Gamma(2m-2k+1)}(-2\sqrt{\pi}) , 0\right).
\end{align*}
Using the bounds in lemma \ref{lemma:alphacbounds},
\begin{align*}
(I)&\leq \sum_{k=1}^m \frac{1}{32 \sqrt{\pi} k^{1/2}} \frac{1}{mk} \frac{1}{\sqrt{2m-1}\sqrt{2k-1}} \leq \frac{1}{32 \sqrt{\pi}},\\
(II)&\leq \frac{1}{4\pi} \sum_{k=1}^m \frac{1}{k^{1/2} (2m-2k+1)^{3/2}} (4 \sqrt{k})\\
&\leq \frac{1}{\pi} \left(1 - \frac{1}{\sqrt{2m-1}}\right) \leq \frac{1}{\pi},\\
(III) &\leq \frac{1}{4\pi} \sum_{k=1}^m \frac{1}{k^{1/2}} \frac{\sqrt{2k-1}}{(2m -2k + 3)^{3/2}}\\
&\leq \frac{1}{\sqrt{24}\pi},\\
(IV) &\leq \frac{1}{4\sqrt{\pi}} \left(\sum_{k=1}^m \frac{(2k)^{1/2}}{k^{1/2}}  \frac{1}{(2m-2k+1)\sqrt{2m-2k}} + \sqrt{\pi}\right)\\
&\leq \frac{1}{2\pi} + \frac{1}{2},\\
(V)&= \frac{1}{4\pi} 4 \pi  \frac{\Gamma(2m+3/2)}{\Gamma(2m+1)}\frac{\Gamma(m+1/2)}{\Gamma(m+1)}\\
&\leq \frac{\sqrt{2m+1}}{\sqrt{m}} \leq \sqrt{3}.
\end{align*}
Finally,
\begin{align*}
\sum_{k=0}^m \frac{\Gamma(k+1/2)}{\Gamma(k)} Q(2m,2k) &\leq (I) + (II) + (III) + (IV) + (V)\\
&\leq 2.8.
\end{align*}
To deduce the inequality (\ref{evenqbound}), we use the previously derived bounds to show that
\begin{align*}
Q(2m-1,2k-1) &\leq \frac{1}{4\pi} \sum_{i=1}^{2k-3} \sqrt{i+1} \frac{1}{(2k-i-2)^{3/2} (2m-i)^{3/2}}\\
&+ \frac{1}{4\pi} \sqrt{\pi} \frac{\Gamma(2k-1/2)}{\Gamma(2k-1)} \frac{\Gamma(2m-2k+1/2)}{\Gamma(2m-2k+2)}\\
&+ \frac{1}{4\pi} \max \left( \frac{\Gamma(2k+1/2)}{\Gamma(2k)} \frac{\Gamma(2m-2k+1/2)}{\Gamma(2m-2k+2)}(-2\sqrt{\pi}) , 0\right),
\end{align*}
so that $Q(2m-1,2k-1) \leq Q(2m,2k)$.  Now it suffices to note that in the full sum,\\
$\sum_{k=1}^m \frac{\Gamma(k+3/2)}{\Gamma(k+1)} Q(2m-1,2k-1) \leq 2 \sum_{k=1}^m \frac{\Gamma(k+1/2)}{\Gamma(k)} Q(2m-1,2k-1)$ and we get
\[
\sum_{k=1}^m \frac{\Gamma(k+3/2)}{\Gamma(k+1)} Q(2m-1,2k-1)\leq 2 \sum_{k=1}^m\frac{\Gamma(k+1/2)}{\Gamma(k)} Q(2m,2k)  \leq 5.6.
\]
$\qed$}

We now return our focus to finding a bound on the expression for $\alpha_{\RR}(d)$ given in (\ref{alpharcdiff}).  Since $\psi_1,\psi_2$ depend on the parity of $d$, we split in to two cases.\\
\\
\textbf{Odd $d=2m+1$}\\
From (see \cite{Gradshteyn_Ryzhik} section 7.414 equation 6),
 \[
 \int_0^{\infty} e^{-x/2} L_{2m}(x) dx = 2,
\]
thus equation  (\ref{alpharcdiff}) becomes
\[
\alpha_{\CC}(2m+1) - \alpha_{\RR}(2m+1) = \frac{1}{(2m+1)^{3/2}} \frac{\Gamma(m+1)}{\Gamma(m+1/2)} \left(\sum_{k=0}^{m} \frac{\Gamma(k+1/2)}{\Gamma(k)} Q(2m,2k) - 2^{1/2}\right),
\]
and using the first bound in Lemma \ref{lemma:qbounds},
\[
\alpha_{\CC}(2m+1) - \alpha_{\RR}(2m+1) \leq 2.8 \sqrt{m+1/2} \frac{1}{(2m+1)^{3/2}} \leq m^{-1}.
\]

\textbf{Even $d=2m$}\\

For $d=2m$, we have
\[
\alpha_{\RR}(2m) = \alpha_{\CC}(2m) - \frac{1}{2(2m)^{3/2}} \int_0^{\infty} x^{1/2} \frac{\Gamma(m+1/2)}{\Gamma(m)} L_{2m-1}(x) \{ \psi_1(x) - \psi_2(x)\} dx.
\]
We split the integral into two parts,
\[
(I) := \frac{1}{2(2m)^{3/2}} \int_0^{\infty} x^{1/2} \frac{\Gamma(m+1/2)}{\Gamma(m)} L_{2m-1}(x)  \psi_1(x) dx 
\]
\[
(II) := \frac{-1}{2(2m)^{3/2}} \int_0^{\infty} x^{1/2} \frac{\Gamma(m+1/2)}{\Gamma(m)} L_{2m-1}(x) \psi_2(x) dx.
\]
Expanding from the definition of $\psi_1$ above, we have

\begin{align*}
(I) &= \frac{1}{2(2m)^{3/2}} \int_0^{\infty} \frac{\Gamma(m+1/2)}{\Gamma(m)} x^{1/2} L_{2m-1}(x)e^{-x} \sum_{k=0}^{m-1}  \frac{\Gamma(k)}{\Gamma(k+1/2)} L_{2k-1}(x) dx\\
&= \frac{1}{2(2m)^{3/2}} \frac{\Gamma(m+1/2)}{\Gamma(m)} \sum_{k=1}^{m} \frac{\Gamma(k+3/2)}{\Gamma(k+1)} Q(2m-1,2k-1),
\end{align*}
so by Lemma \ref{lemma:qbounds},
\begin{align*}
(I) & \leq \frac{1}{2(2m)^{3/2}} \frac{\Gamma(m+1/2)}{\Gamma(m)} 5.6  \leq \frac{1}{ m^{1/2}}.
\end{align*}
The other part of the integral is
\begin{align*}
(II) &= \frac{-1}{2(2m)^{3/2}} \int_0^{\infty} x^{1/2} \frac{\Gamma(m+1/2)}{\Gamma(m)} L_{2m-1}(x) (x/2)^{-1/2} e^{-x/2} \left[\frac{2 \Gamma(1/2,x/2)}{\Gamma(1/2)} - 1\right] dx\\
&= \frac{-1}{4m^{1/2}} \int_0^{\infty} \frac{\Gamma(m+1/2)}{\Gamma(m)} L_{2m-1}(x) e^{-x/2} \frac{2 \Gamma(1/2,x/2)}{\Gamma(1/2)} dx + \frac{1}{2m^{3/2}} \frac{\Gamma(m+1/2)}{\Gamma(m)},
\end{align*}
where we use the fact that for odd $2m-1$ (see \cite{Gradshteyn_Ryzhik} section 7.414 equation 6), 
\[
\int_0^{\infty} L_{2m-1}(x) e^{-x/2} dx = -2.
\]
We can bound the first integral in the expression of (II) by
\begin{align*}
\left| \int_0^{\infty} L_{2m-1}(x) e^{-x/2} \Gamma(1/2,x/2) dx\right|  &\leq \left(\int_0^{\infty} e^{-x} L_{2m-1}(x)^2 dx\right)^{1/2} \left(\int_0^{\infty} \Gamma(1/2,x/2)^2 dx\right)^{1/2}\\
&= \left[\int_0^{\infty}  \left(\int_x^{\infty} t^{-1/2} e^{-t} dt\right)^2 dx\right]^{1/2}\\
&= \left[\int_0^1 \left(\int_x^{\infty} t^{-1/2} e^{-t} dt \right)^2 dx + \int_1^{\infty} \left(\int_x^{\infty} t^{-1/2} e^{-t} dt\right)^2 dx\right]^{1/2}\\
&\leq\left (\Gamma(1/2)^2 + \int_1^{\infty} (e^{-x})^2 dx\right)^{1/2} \leq(\pi + e^{-2}/2)^{1/2},
\end{align*}
so finally
\begin{align*} (II) &\leq \frac{ (\pi + 1/2 e^{-2})^{1/2}}{2\sqrt{\pi} m^{3/2}} \frac{\Gamma(m+1/2)}{\Gamma(m)} + \frac{m^{1/2}}{2m^{3/2}}\\
 &\leq 1.01 m^{-1}.
\end{align*}

Combining the above bounds we see that in the case of even $d=2m$,
\[
\alpha_{\CC}(2m) - \alpha_{\RR}(2m) =(I) + (II) \leq 2.01 m^{-1}.
\]
$\qed$
}





\end{document}